\documentclass[12pt]{iopart}
\usepackage{iopams}

\usepackage[lofdepth,lotdepth]{subfig}

\usepackage[pdftex]{graphicx}
\usepackage{epstopdf}
\usepackage{algorithm}
\usepackage{algpseudocode}
\usepackage{color}
\usepackage{cite}
\usepackage[cmex10]{amsmath}
\usepackage[cmex10]{amsmath}
\usepackage{amssymb}
\usepackage{gensymb}
\usepackage{hyperref}
\usepackage{multicol}
\usepackage[T1]{fontenc}
\usepackage{graphicx}
\usepackage{subfig}
\usepackage{siunitx}
\usepackage{breqn}
\usepackage{algpseudocode}
\usepackage{lipsum}
\usepackage{cuted}
\usepackage{amsmath}

\newcommand{\R}[1]{{\color{black}#1}}

\newcommand{\dif}{{\rm d}}

\newcommand{\vJ}{{\bf J}}
\newcommand{\vE}{{\bf E}}

\newcommand{\vA}{{\bf A}}

\begin{document}
\title[Thermal quench modeling of REBCO racetrack coils ...]{\R{Thermal quench modeling of} REBCO racetrack coils under conduction cooling at 30 K for aircraft electric propulsion motors}

\author{Arif Hussain, Anang Dadhich \& Enric Pardo\footnote{\R{Corresponding author.}}}

\address{Institute of Electrical Engineering, Slovak Academy of Sciences, Dubravska Cesta 9, 84101, Bratislava, Slovakia }
\ead{enric.pardo@savba.sk}
\vspace{10pt}
\begin{indented}
\item[]May 2025
\end{indented}

\begin{abstract} \\

High-temperature superconducting (HTS) racetrack coils are promising components for lightweight, high-power electric machines due to their exceptional current-carrying capacity. However, self-heating due to AC loss or DC short circuits can cause electro-thermal quench, which poses a significant challenge for the design and reliability of superconducting motors. In this study, we apply a novel computational approach that integrates the Minimum Electro-Magnetic Entropy Production (MEMEP) method for precise electromagnetic calculations with the Finite Difference Method (FDM) for detailed electrothermal analyses. This coupled approach enables simultaneous calculations to capture the dynamic interplay between electromagnetic and thermal behaviors. The investigation focuses on the response of an HTS racetrack coil subjected to short-circuit DC voltages ranging from low (1 V) to high values (1000 V) at a cryogenic operating temperature of 30 K. The simulations were conducted under two thermal boundary conditions: complete adiabatic conditions and cooling applied to one side of the coil with the top surface fixed at 30 K. The results reveal distinct current and temperature dynamics when varying the voltage amplitude. At higher voltages, the current exceeds the critical value, causing rapid thermal runaway that damages the superconducting material. In contrast, at lower voltages, the coil presents periodic oscillations in current and temperature, demonstrating a complex interplay of thermal diffusion and electromagnetic stability. This study provides critical insights into the thermal management and fault response of HTS coils for aviation applications, particularly in the design of superconducting motors for electric aircraft. By understanding the thermal stabilization mechanisms and current dynamics under varying fault conditions, this work contributes to the development of efficient, reliable, and environmentally sustainable propulsion systems aimed at reducing carbon emissions in aviation.

\end{abstract}

\section{Introduction}

Superconducting motors have gained significant attention for aviation applications due to their potential for high power density, efficiency, and reduced weight compared to conventional electrical machines \cite{haranKS2017SST, Yingnan_2023}. Through the use of superconductors, these motors can achieve nearly lossless conduction, enabling improved performance in high-power aircraft propulsion systems \cite{ASCEND2022JCS}. However, the complex interplay between electromagnetic and electrothermal \R{behavior} in superconducting windings, particularly under dynamic operating conditions, remains a critical area of study. Understanding \R{this behavior} is essential for optimizing the reliability and efficiency of superconducting racetrack coils in aviation motors \cite{Yuan_2025}. Superconducting electric propulsion motors are especially interesting for hydrogen-electric aircraft, since liquid hydrogen can serve as both fuel (possibly in fuel cells) and coolant \cite{ASCEND2022JCS, airbus_upnext,aerospace2024}.

The study of electromagnetic and electrothermal behavior in superconducting racetrack coils has the potential to significantly impact the development of advanced superconducting technologies. Enhancing the understanding of these complex interactions can lead to improved designs that optimize the efficiency, reliability, and performance of superconducting systems. These advancements are crucial not only for the aerospace industry but also for a wide range of applications, including high-field magnets, high-speed transportation, and power generation \cite{Dezhin2022, gaoY2023PhC}. The insights gained could accelerate the adoption of superconducting technologies, contributing to the advancement of more efficient and sustainable solutions in various sectors.

Superconducting motors for hydrogen-electric aircraft \R{present} challenges, including the need for efficient cryogenic cooling at around 30 K, quench detection, and rapid thermal management. The operating temperature of the superconducting windings needs to be higher than liquid hydrogen (20 K) in order to form a temperature gradient, and hence enable cooling. That is why we can set a goal of around 30 K of operation temperature, which leaves a safe margin \cite{aerospace2024}. Moderate magnetic fields help avoiding high mechanical strains (in contrast to high-field magnets, where strains are high), and optimizing coil impregnation ensures long-term durability \cite{aerospace2024}. Superconducting racetrack coils in motors contribute to achieve the high power-to-weight ratios (approximately 25 kW/kg) required for zero-emission airliners \cite{FlyZero2022roadmap}. REBCO ($RE$Ba$_2$Cu$_3$O$_{x-7}$ where $RE$ \R{is a} rare-earth, typically Y, Gd or Sm) is especially advantageous in rotor designs, enhancing magnetic fields and power density compared to materials like magnesium diboride (MgB$_2$). REBCO retains high engineering current densities at elevated temperatures, such as 30 or 40 K, whereas MgB$_2$ experiences a significant drop in engineering critical current density just a few degrees above 20 K. \R{Then,} REBCO enables efficient operation at 20-30 K, \R{or even higher temperatures} \cite{ASCEND2022JCS, kalsiSS2023IES}. Advanced designs further minimize AC losses and provide magnetic shielding through stacking effects \cite{pardoE2019IES, pardoE2019EUCAS}.

Advanced numerical modeling plays a crucial role in studying the electromagnetic and electrothermal behavior of superconducting racetrack coils, helping to predict electromagnetic fields, current density distributions, and thermal profiles. Recent improvements in electrothermal simulations have enhanced coil design, cooling strategies, and material selection, contributing to the optimization of superconducting motors and generators \cite{Yutaka2021}.

Advanced numerical approaches provide a powerful tool to tackle the challenges in designing superconducting \R{power applications} \cite{dutoitB}. These techniques allow for in-depth studies of the electromagnetic and electrothermal behavior within superconducting racetrack coils, giving us a better understanding of how currents, temperatures, and power loss interact with each other \cite{pardoE2019IES, benkelT2020IES, santos_2022, haroE2013IES}. By accurately simulating the distribution of electromagnetic fields, current density, and thermal effects, these models help pinpoint areas for improvement in coil design, cooling systems, and material choices. They also make it possible to predict potential issues like quenching or thermal hotspots, allowing for better cooling strategies and more efficient thermal management \cite{fazilleauP2024IES, dadhichA2024SST}. With these simulations, we can refine the design process, ensuring \R{that} superconducting motors perform efficiently, remain durable, and meet the demands of cryogenic environments. \R{Other notable works on single tapes are \cite{royF2008IES, bonnardCH2017SST}.}

This study investigates superconducting racetrack coils under both adiabatic conditions and cooling from the top surface at 30 K, powered by voltage sources, to more accurately represent motor applications than current-driven configurations. Alternating voltages simulate standard stator operation, while DC voltages model fault conditions, such as inverter malfunctions or short-circuits due to insulation damage. Given the low resistance of superconducting coils, they are highly sensitive to DC faults, which can trigger electrothermal quenches.

For this study, we developed a custom numerical model combining the Minimum Electro-Magnet Entropy Production (MEMEP) method for electromagnetic calculations and the Finite Difference method (FDM) for electrothermal calculations. The MEMEP method, as described in previous works \cite{pardoE2015SST, pardoE2017JCP}, is coupled with FDM to provide efficient thermal simulations, as discussed in details in our recent studies \cite{Hussain_2024, dadhichA2024SST}. Initial tests at 77 K revealed unexpected behavior, where moderate DC voltages caused non-uniform electrothermal quenches due to AC losses \cite{Hussain_2024}. This work extends the analysis to 30 K operation, with the critical temperature ($T_c$) set at 92 K. Our model is adaptable to low temperatures, such as those used in hydrogen-electric aircraft, and provides valuable insights into the electrothermal dynamics of superconducting coils in realistic motor conditions. Preliminary findings were presented at conferences like in \cite{hussainA2025CCA}.

\section{Studied Configuration}

\label{s.studied}
 
\begin{figure}[htp]
\centering
\subfloat[]{
  \includegraphics[trim=0 0 0 0,clip,width=6 cm]{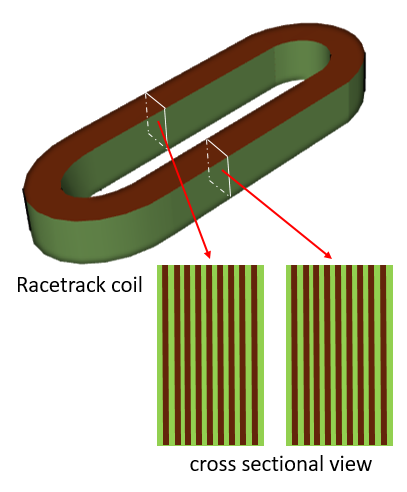}
}
\subfloat[]{
  \includegraphics[trim=0 0 0 0,clip,width=8 cm]{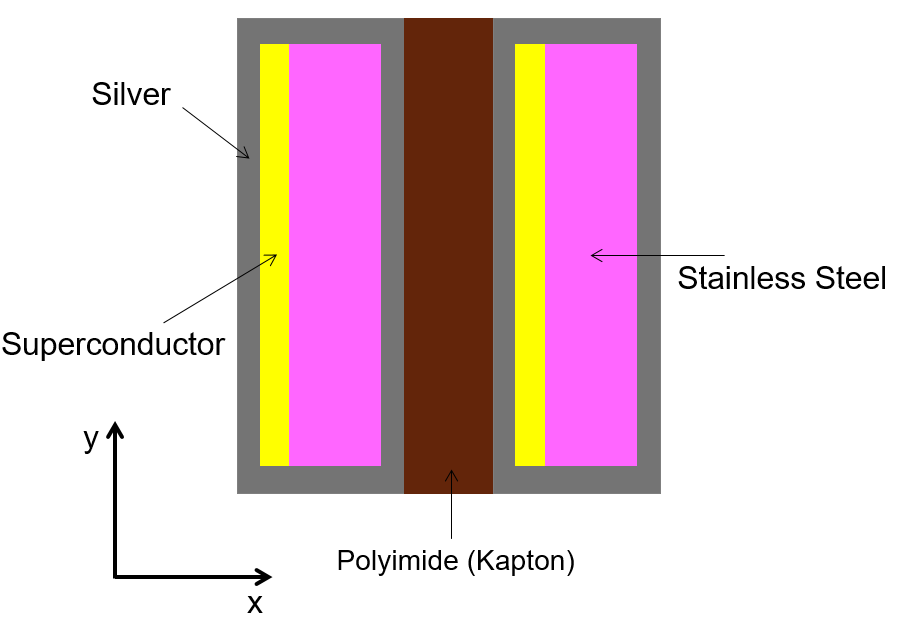}
}
\caption{(a) {Sketch} of a racetrack coil and its cross-section, where the green part represents the superconducting tape with all its layers and the red part is the electric insulation. (b) Cross-sectional view of two neighboring superconducting turns with polyimide (Kapton) insulation in between.}
\label{f.coil_and_tape}
\end{figure}

In this article, we study the configuration of the racetrack coil from REBCO superconducting tape (figure \ref{f.coil_and_tape}(a)) as already discussed in our recent study \cite{Hussain_2024} but submitted to conduction cooling with a cold sink at 30 K, instead of liquid nitrogen bath. In this study, we examine the multi-layer of a superconducting tape, incorporating silver, superconductor, hastelloy, and polyimide (Kapton) as an electrical insulator with the thicknesses of 4 $\mu$m, 2 $\mu$m, 100 $\mu$m, and 20 $\mu$m respectively (these are total thicknesses across the whole tape). Silver has an exceptional electrical and thermal conductivity, which aids in stabilizing the tape, dissipating heat, while hastelloy as a substrate brings high heat capacity and incredible strength for superconducting material (Figure \ref{f.coil_and_tape}(b)). In this article, we assume that the thermal properties of Hastelloy are the same as stainless steel because of the availability of measured data of the heat conductivity and capacity as a function of temperature, $k(T)$ and $C_v(T)$ \cite{NIST_database}.

In this study, we model the electrical behavior of the superconducting layer using a power-law relationship between the electric field $\vE$ and the current density of the macroscopic super-currents, $\vJ_s$, as follows \cite{grilliF2014IES, pardoE2023book}:
\begin{equation}
\vE(\vJ_s) = E_c \left(\frac{|\vJ_s|}{J_c}\right)^n \frac{\vJ_s}{|\vJ_s|},
\label{e.EJ1}
\end{equation}
where $E_c$ = $10^{-4}$ V/m is the critical current criterion, $J_c$ represents the critical current density, and $n$ is the power-law exponent. The exponent $n$ defines the smooth transition of the $\vE(\vJ_s)$ curve, from negligible electric fields $E\approx 0$ when $|\vJ_s|<J_c$, to rapidly growing electric fields with $|{\bf J}_s|$ when $|\vJ_s|>J_c$. While more complex models for the $\vE(\vJ_s)$ relation exist and can be incorporated in our model \cite{siroisF2019IES, rivaN2021SST}, the power-law approach provides a straightforward and widely used framework for analyzing the superconductor's response.

In this study, we assume constant critical current ($I_c$) for simplicity. This value is determined using the load-line technique with the measured tape $I_c$ in the database in \cite{robinson_data} as input. This is a standard method to estimate the operating point of superconducting coils. Using this method, and the tape $I_c$ data for SuperOx YBCO 2G HTS, we obtain a critical current of 470 A at 30 K, which corresponds to $J_c=5.785\cdot 10^{10}$ Am$^{-2}$.

\section{Numerical Modeling Methods}

This section introduces the electromagnetic model, thermal model, and their interaction developed for the computations in this study. These self-programmed methods are computationally efficient. Notably, the thermal model employs an innovative finite-difference method (FDM) approach, allowing for temperature-dependent thermal properties and material discontinuities. This model is seamlessly coupled with a fast and accurate electromagnetic model based on variational principles.

\subsection{Electromagnetic Model: Variational Principle}

This study employs a numerical modeling approach based on the Minimum Electro-Magnetic Entropy Production (MEMEP) principle to simulate the electromagnetic behavior of racetrack coils in superconductors \cite{pardoE2017JCP}. This approach is based on variational formulation, where the electric field $\vE$ is expressed in terms of the vector potential $\vA$, scalar potential $\phi$, and current density ${\bf J}$ in 
\begin{equation}
{\vE (\vJ) = - {\frac{\partial \vA}{\partial t}} - {\triangledown\phi}}.
\label{e.EArelation}
\end{equation}
The model accounts for any non-linear ${\bf E}({\bf J})$ relation \cite{pardoE2017JCP}. This relation can also be non-homogeneous due to temperature-dependent properties, such as the critical current density. For infinitely long configurations, the problem is reduced to a 2D cross-sectional model \cite{pardoE2023book}, further simplifying computations (figure \ref{f.coil_and_tape}). The current density $J$ is solved under either current or voltage constraints, with the later applied in this work. Due to symmetry in the infinitely long z-direction, the electric field and current density can be expressed as $\vE=E{\bf \hat{z}}$ and $\vJ=J{\bf \hat{z}}$, respectively. Within the framework of our numerical modeling approach, solving (\ref{e.EArelation}) is achieved by minimizing the functional in (\ref{e.L2D}) at every time step, which presents a unique minimum \cite{Hussain_2024}. This functional forms the basis for the variational principles employed in our simulations.

\begin{equation} 
L[\Delta J]=\int_S {\rm d}s \left [ \frac{1}{2} \Delta J\frac{\Delta A}{\Delta t} + \Delta J \frac{\Delta A_a}{\Delta t} + U(J) \right ] -VI. 
\label{e.L2D}
\end{equation}

where, $\Delta t$ is a certain time step, $\vJ$ is the current density at time $t$, $\Delta \vJ=\vJ(t)-\vJ(t-\Delta t)$, and we assume that $\vJ(t-\Delta t)$ is known. For the zero-field cooling case, $\vJ=0$ at $t=0$. Then, we can compute the whole time evolution by applying this method to subsequent time steps. In the equation above, $U (\vJ)$ is defined as the dissipation factor, which is
\begin{equation}
{ U ( \vJ) = \int_{0}^{ \vJ} \dif {\bf J}' \cdot \vE({\bf J}')}.
\label{e.disspiation}
\end{equation}
The detailed discussion can be found in our previous studies \cite{pardoE2023IES, pardoE2023book, pardoE2015SST}.

In this article we employ a simplified homogeneous approach for superconducting tapes, bypassing the intricate multi-layer geometry of the tapes. The nonlinear resistivity of the superconducting layer is described by the power-law $E(J_s)$ relation of (\ref{e.EJ1}), and a constant power-law exponent, $n=30$. The parallel connection of superconducting and metallic layers is modeled. To simplify computations, the effective resistivity for the entire tape is approximated as detailed in \cite{pardoE2023IES}. The approach has been benchmarked against experimental results \cite{pardoE2015SST} and other numerical methods, demonstrating high accuracy and computational efficiency.

\subsection{Thermal Model}

Superconductors exhibit time-dependent electromagnetic properties that can induce dissipation, leading to significant temperature rise. We begin with the general heat diffusion equation:
\begin{equation}
{ C_{pv} (T) \frac{{\partial}T}{{\partial}t}= \nabla \cdot (\bar{\bar k}(T)\cdot \nabla T) + p(\vJ)},
\label{e.FD}
\end{equation}
where $C_{pv}$ is the heat capacity per unit volume at constant pressure, $p(\vJ)$ is the power heat density generated by electromagnetic dissipation, which is $p(\vJ)={\bf E}({\bf J})\cdot{\bf J}$, and $\bar{\bar k}$ is the heat capacity tensor that accounts for the material anisotropy. Equation (\ref{e.FD}) describes how thermal energy is transferred through a medium over time. This equation is also valid when $\bar{\bar k}$ and $C_{pv}$ are non-homogeneous. We can discretize (\ref{e.FD}) as detailed in \cite{Hussain_2024} and solve $T$ by the finite difference method. This discretization takes nonhomogeneous heat conductivity and capacity into account. We show temperature dependent heat conductivity $k(T)$ and specific heat $C_{pv}(T)$ of different materials used in our model in table \ref{t.table_1}, based on the database in \cite{NIST_database}. In this work, we discretize the diffusion equation in 2D using the finite difference scheme detailed in \cite{Hussain_2024}.

\begin{table}[tpb]
\begin{center}
\begin{tabular}{|*{7}{c|}}
\hline
{\textbf{Temperature}}  & \multicolumn{2}{c|}{\textbf{Silver}}
                            & \multicolumn{2}{c|}{\textbf{Stainless Steel}} 
                            & \multicolumn{2}{c|}{\textbf{Polyimide (Kapton)}}\\
\cline{2-7}
 [K]  &  {k} & {$C_{pv}$} & {k}& {$C_{pv}$} & {k} & {$C_{pv}$} \\
                        & [W/m.K] & [$J/(kg\cdot K)$]  &  [W/m.K] & [$J/(kg\cdot K)$] & [W/m.K] & [$J/(kg\cdot K$)]\\
\hline
{30} & {1930} & {54.392} & {3.468} & {30.539} & {0.067} & {116.229}\\
\hline
{40} & {1050} & {89.161} & {4.670} & {58.687} & {0.083} & {172.862}\\
\hline
{50} & {700} & {118.365} & {5.730} & {95.690} & {0.097} & {224.323}\\
\hline
{60} & {550} & {139.745} & {6.646} & {136.915} & {0.108} & {270.398}\\
\hline
{80} & {471} & {170.246} & {8.114} & {215.259} & {0.128} & {348.942}\\
\hline
{100} & {450} & {190.623} & {9.223} & {275.496} & {0.141} & {413.687}\\
\hline
{200} & {430} & {225.726} & {12.632} & {416.429} & {0.174} & {626.773}\\
\hline
{300} & {429} & {237} & {15.308} & {469.449} & {0.192} & {754.550}\\
\hline
\end{tabular}
\caption{Thermal properties of the materials in the superconducting tape for our model.}
\label{t.table_1}
\end{center}
\end{table}

\begin{figure}[htp]
\centering
\subfloat[]{
  \includegraphics[trim=0 0 0 0,clip,width=8 cm]{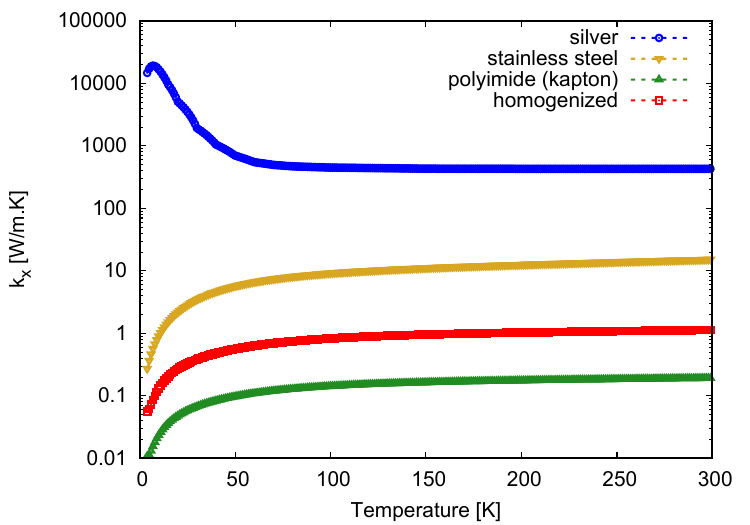}
}
\subfloat[]{
\includegraphics[trim=0 0 0 0,clip,width=8 cm]{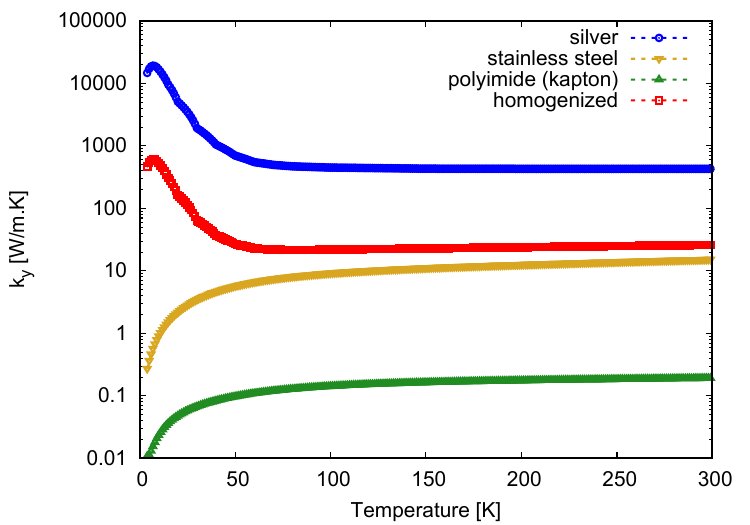}
}
\caption{ Sketch of (a) temperature dependent thermal conductivity of each material in x-direction $k_x(T)$. (b) temperature dependent thermal conductivity of each material in y-direction $k_y(T)$.}
\label{f.cond_heat}
\end{figure}

\begin{figure}[htp]
\centering
\includegraphics[trim=0 0 0 0,clip,width=8 cm]{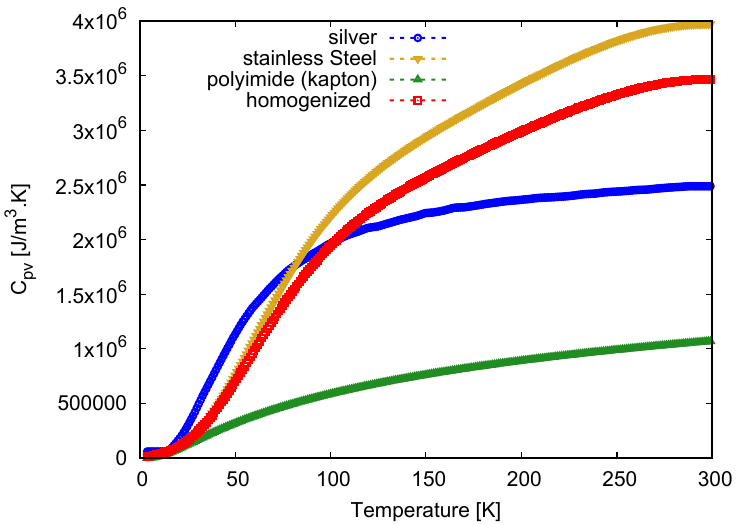}

\caption{ Sketch of temperature dependent specific heat $C_{pv}(T)$ of each}
\label{f.spec_heat}
\end{figure}

\section{Homogenization Approach for Electromagnetic and Electrothermal Models}

In this study, we adopt a homogenized approach for both the electromagnetic and electrothermal properties of the superconducting tape, referring to our prior work for detailed descriptions \cite{Hussain_2024, pardoE2023IES, dadhichA2024SST}. For the electromagnetic analysis, we utilize the homogeneous model of the superconducting tape, treating the layered structure as an effective medium with equivalent resistivity, incorporating both the superconducting and metallic components. The $E(J_s)$ power-law relation defines the non-linear resistivity of the superconductor. We estimate the equivalent non-linear resistivity as the parallel connection of the superconductor and all metal layers, following \cite{pardoE2023IES}. Figure \ref{f.cond_heat} illustrates the temperature dependent homogenized thermal conductivity along the thickness $k_x(T)$ (figure \ref{f.cond_heat}a), homogenized thermal conductivity along the width $k_y(T)$ (figure \ref{f.cond_heat}b), and homogenized specific heat $C_{pv}(T)$ (figure \ref{f.spec_heat}) of the materials used in our racetrack coil. These plots help visualize the overall trends in material behavior. For precise numerical values, please refer to table \ref{t.table_1}, which provides detailed temperature-dependent data for each material.

For the thermal analysis, we apply a homogenized model for the thermal properties of the tape, accounting for layers with temperature dependent heat capacities per unit volume, $C_{pv}$, and thermal conductivities $k(T)$ along the $x$ and $y$ directions. The effective thermal properties, such as the volumetric heat capacity and thermal conductivities ($C_{pv,e}$, $k_{x,e}$, and $k_{y,e}$), are computed using weighted averages and integrals over the tape geometry, ensuring accurate representation of the layered structure within the model \cite{Hussain_2024}. This simplified approach enables efficient simulation while retaining the essential physical characteristics of the tape.

\section {Results and Discussion}

In this paper, we analyze the electromagnetic and thermal response of a superconducting racetrack coil under short-circuit DC voltage of several magnitudes (1000, 100, 10, 1 V). We consider both adiabatic and conduction cooling conditions to capture the coil's response under realistic operational scenarios, providing insights into its electrothermal stability and performance limits.

\subsection{Adiabatic Conditions}

\begin{figure}[htp]
\centering
\includegraphics[trim=0 0 0 0,clip,width=10 cm]{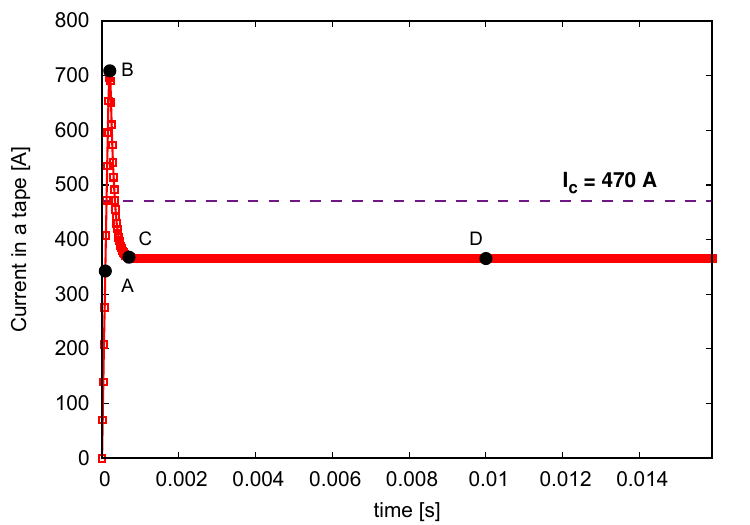}\\
\caption{Total current in racetrack coil for 1000 V DC input with no heat exchange with its surroundings.}
\label{f.1000V_cur}
\end{figure}

\begin{figure}[htp]
\centering
  \includegraphics[trim=0 0 0 0,clip,width=10 cm]{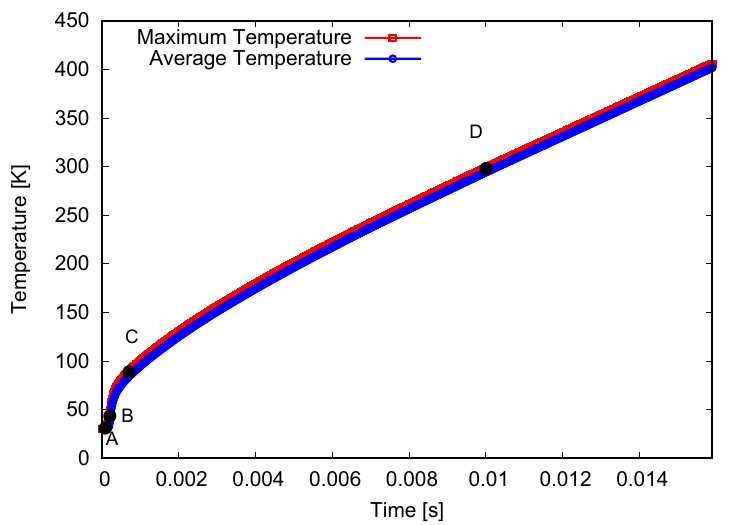} \\
\caption{Temperature rise in racetrack coil for 1000 V DC input with no heat exchange with the cryogenic liquid. } 
\label{f.1000V_temp}
\end{figure}

We start with the case of short-circuit DC of 1000 V with no heat exchange involved. The results show that the maximum and average temperature in the coil rises sharply (figure \ref{f.1000V_cur} and \ref{f.1000V_temp}). Since the current rapidly overcomes the critical current $I_c$, the coil temperature rises sharply, reaching $400$ K within less than $16$ ms (figure \ref{f.1000V_cur} and \ref{f.1000V_temp}). This rapid temperature rise leads to irreversible damage in the superconducting material in very short span of $16$ ms, since above $400$ K the REBCO layer could experience irreversible degradation due to oxygen diffusion out of the superconductor. These findings emphasize the need for robust cooling and current-limiting strategies in superconducting systems to prevent damage under high-fault conditions.
 
\begin{figure}[htp]
\subfloat[initial ramp]{
  {\includegraphics[trim=1.6cm 0 1.4cm 0,clip,height=4.0cm]{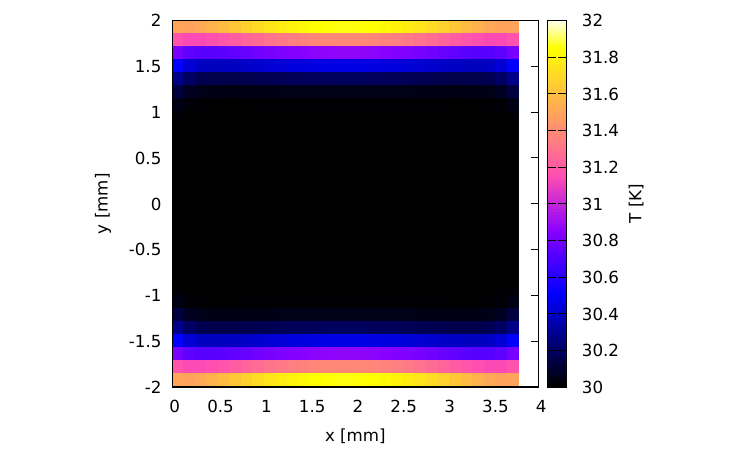}}
}
\subfloat[initial ramp]{
  {\includegraphics[trim=1.6cm 0 1.9cm 0,clip,height=4.0cm]{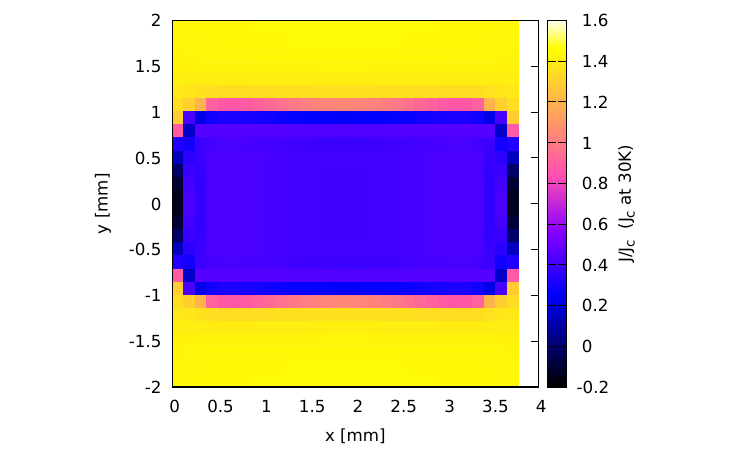}}
}
\subfloat[initial ramp]{
  {\includegraphics[trim=1.6cm 0 1.4cm 0,clip,height=4.0cm]{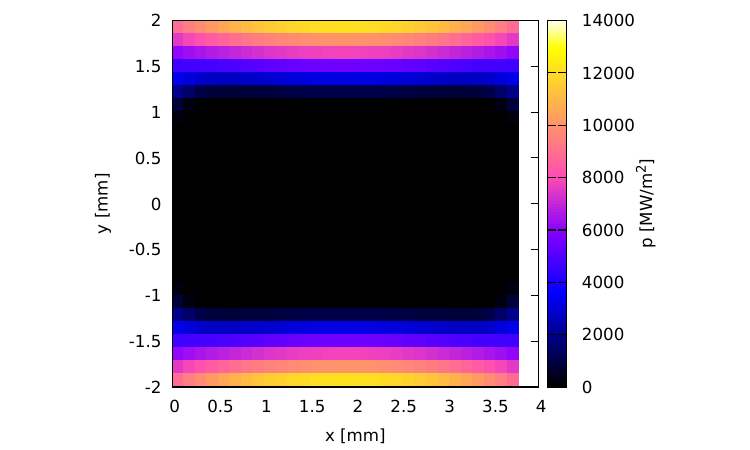}}
}

\hspace{0mm}
\subfloat[current peak]{
  {\includegraphics[trim=1.6cm 0 1.4cm 0,clip,height=4.0cm]{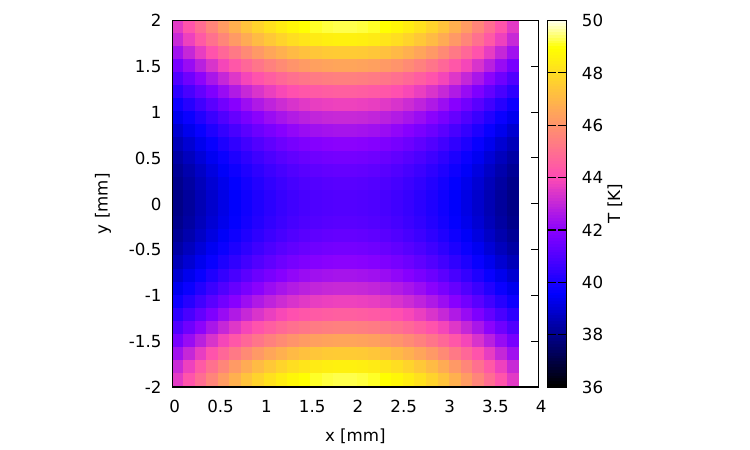}}
}
\subfloat[current peak]{
  {\includegraphics[trim=1.6cm 0 1.4cm 0,clip,height=4.0cm]{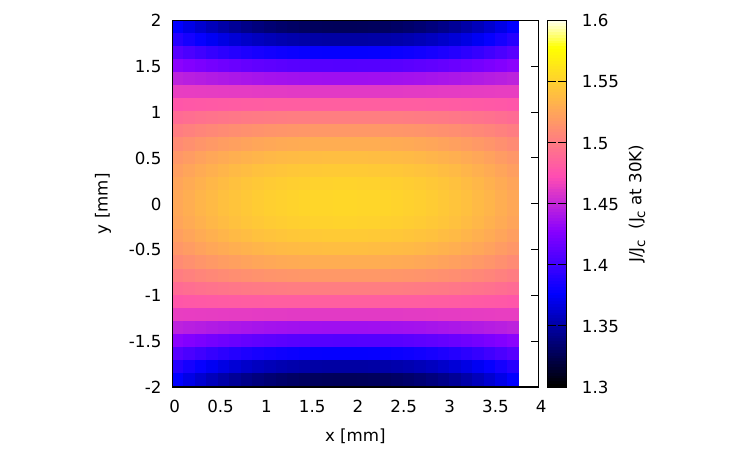}}
}
\subfloat[current peak]{
  {\includegraphics[trim=1.6cm 0 1.4cm 0,clip,height=4.0cm]{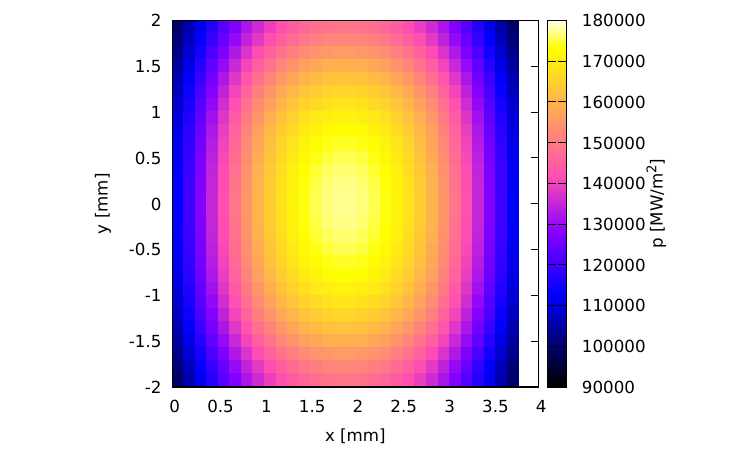}}
}

\hspace{0mm}
\subfloat[first drop]{
  {\includegraphics[trim=1.6cm 0 1.4cm 0,clip,height=4.0cm]{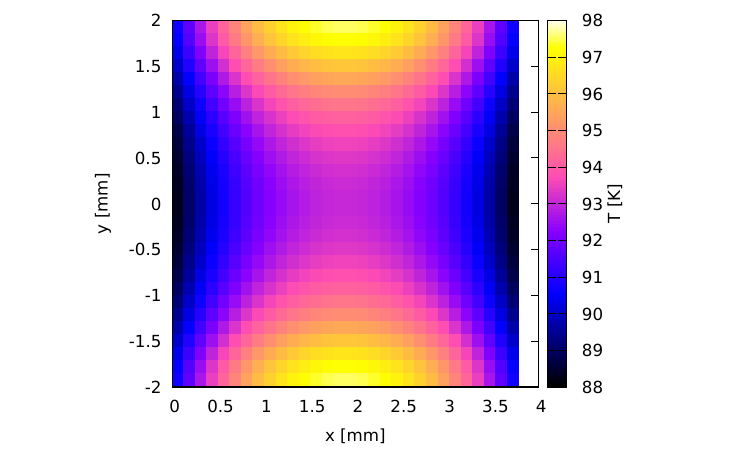}}
}
\subfloat[first drop]{
  {\includegraphics[trim=1.6cm 0 1.4cm 0,clip,height=4.0cm]{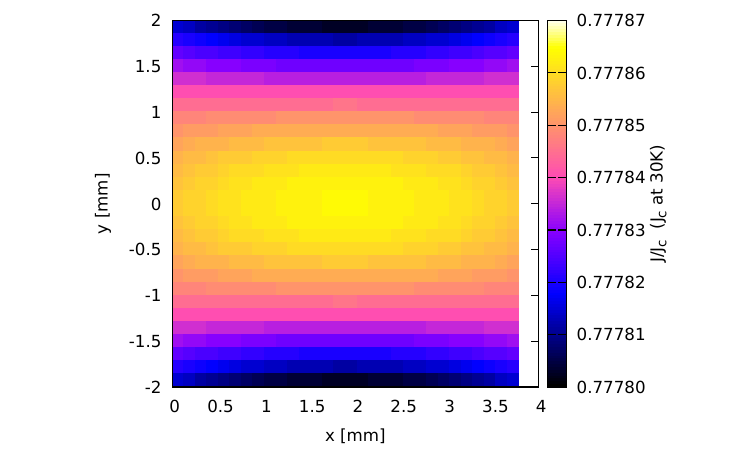}}
}
\subfloat[first drop]{
  {\includegraphics[trim=1.6cm 0 1.4cm 0,clip,height=4.0cm]{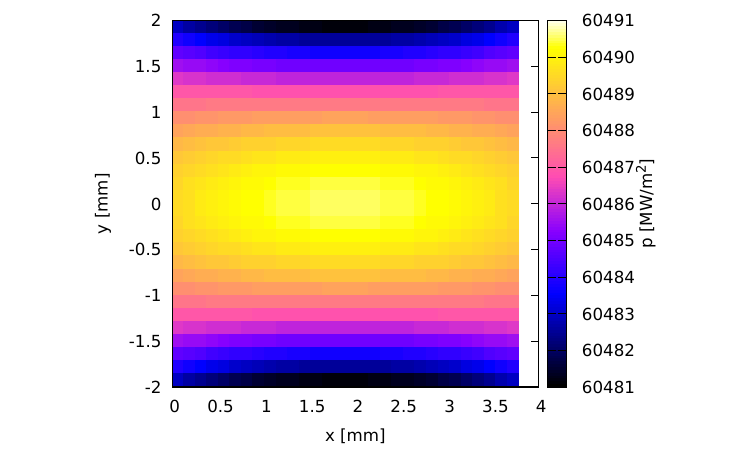}}
}

\hspace{0mm}
\subfloat[stable current]{
  {\includegraphics[trim=1.6cm 0 1.4cm 0,clip,height=4.0cm]{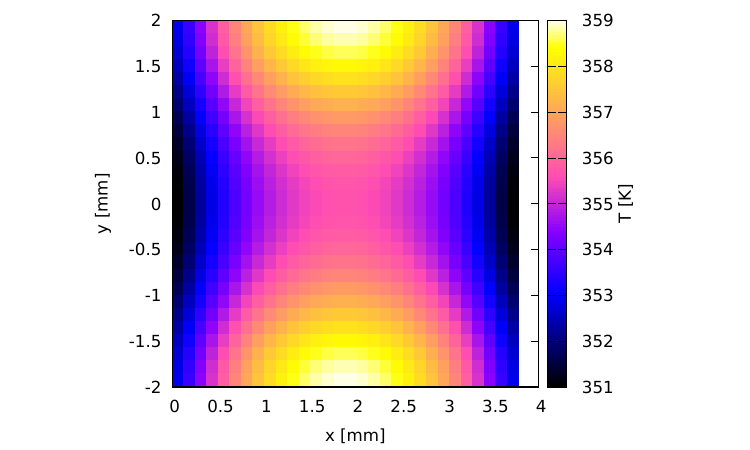}}
}
\subfloat[stable current]{
  {\includegraphics[trim=1.6cm 0 1.4cm 0,clip,height=4.0cm]{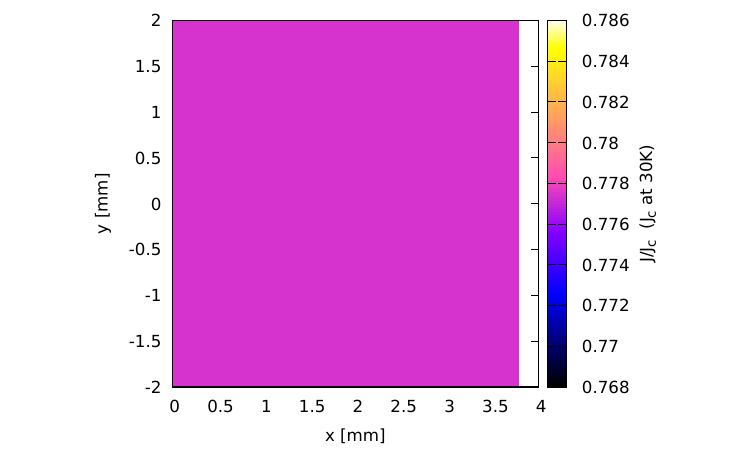}}
}
\subfloat[stable current]{
  {\includegraphics[trim=1.6cm 0 1.4cm 0,clip,height=4.0cm]{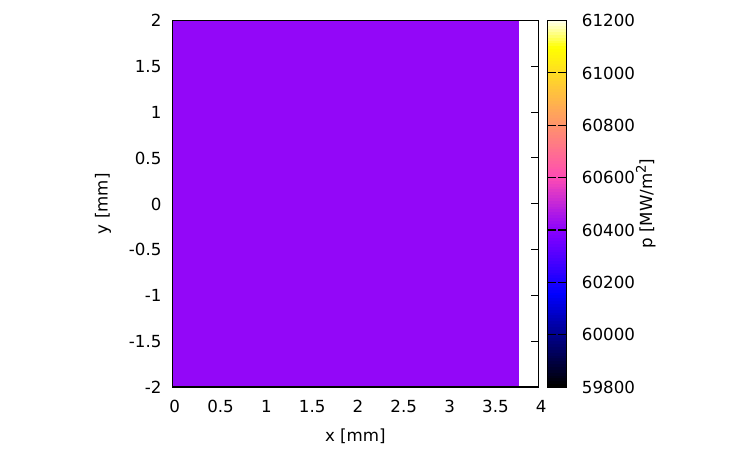}}
}

\caption{Coil cross-section showing maps of temperature (left), current density (middle), and power loss density (right) for adiabatic conditions at 1000 V amplitude of DC voltage for different times: (a,b,c) at point A, (d,e,f) at point B, (g,h,i) at point C, and (j,k,l) at point D in figures \ref{f.1000V_cur} and \ref{f.1000V_temp}.}
\label{f.maps_silver_1000Vdc}
\end{figure}

Next, we analyze the temperature rise, current density, and power loss profiles. For the initial current ramp (point A in figures \ref{f.1000V_cur} and \ref{f.1000V_temp}), we see that temperature starts developing in the racetrack coil on the top and bottom edges due to screening currents, and the corresponding power loss (figure \ref{f.maps_silver_1000Vdc} (a,b,c)). At the current peak, at $2.0 \times 10^{-4}$ s (point B in figures \ref{f.1000V_cur} and \ref{f.1000V_temp}), we observe sudden temperature rise, overcritical state ($|J|>J_c$), and corresponding high power losses (figure \ref{f.maps_silver_1000Vdc} (d,e,f)). Additionally, after the drop in current (point C in figures \ref{f.1000V_cur} and \ref{f.1000V_temp}), the current remains high at around $380$ A, which causes continuous temperature rise and triggers quench ($T > T{_c}$ = 92 K)) and respective high power loss in the racetrack coil (figure \ref{f.maps_silver_1000Vdc} (g,h,i)). At $10$ ms (point D in figures \ref{f.1000V_cur} and \ref{f.1000V_temp}), the whole coil has uniform current density, power loss and almost uniform temperature (figure \ref{f.maps_silver_1000Vdc}(j,k,l)). The current density and power loss are uniform because the whole cross-section is at normal state and we assume temperature-independent resistivity of the metals.

\begin{figure}[htp]
\centering
\includegraphics[trim=0 0 0 0,clip,width=9 cm]{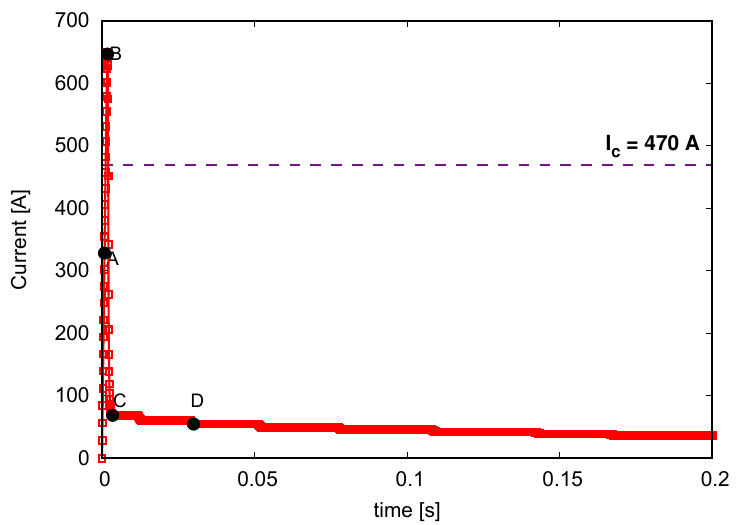}
\\
\includegraphics[trim=0 0 0 0,clip,width=9 cm]{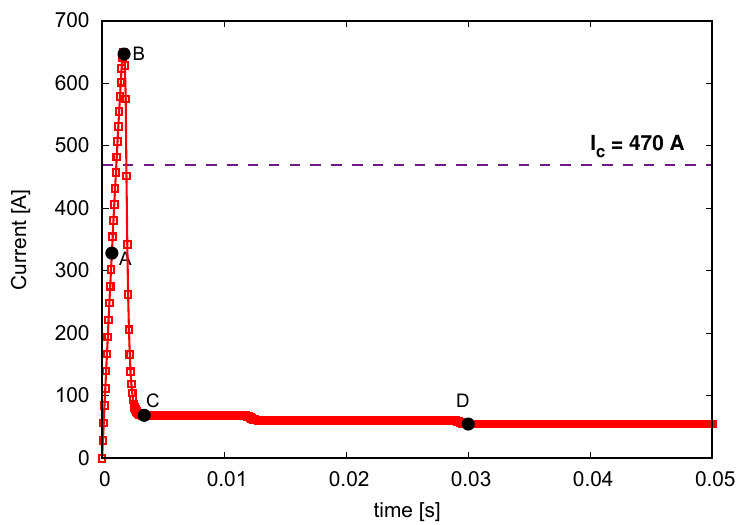}
\caption{(a) Total current in racetrack coil for 100 V DC input with no heat exchange with its surroundings. (b) Zoomed sketch of total current from $0$ to $50$ ms.}
\label{f.100V_cur}
\end{figure}

\begin{figure}[htp]
\centering
  \includegraphics[trim=0 0 0 0,clip,width=9 cm]{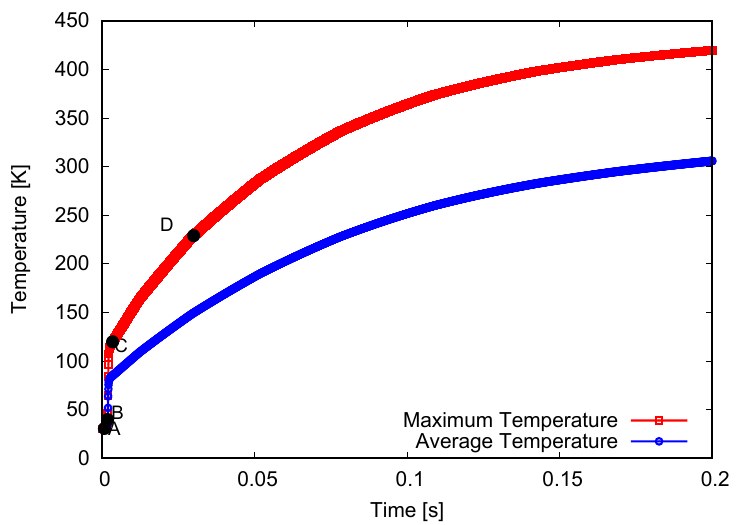} \\
  \includegraphics[trim=0 0 0 0,clip,width=9 cm]{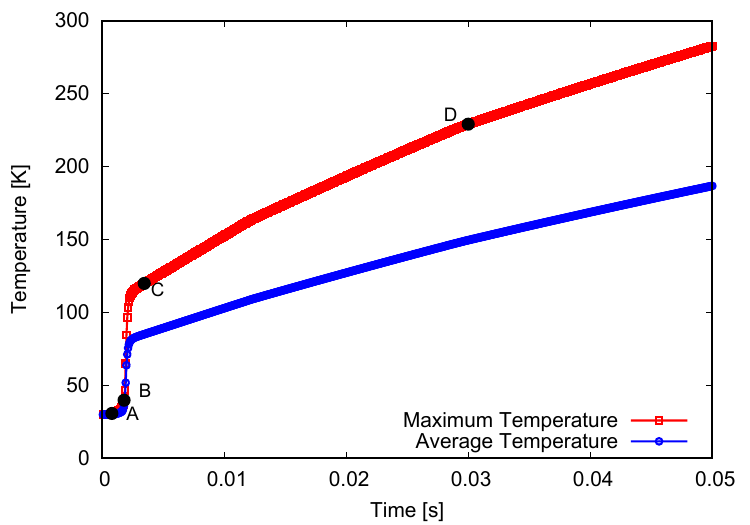}
\caption{ (a) Temperature rise in racetrack coil for 100 V DC input with no heat exchange with the cryogenic liquid. (b) Zoomed sketch of temperature from $0$ to $50$ ms.} 
\label{f.100V_temp}
\end{figure}

In a similar manner, when a short-circuit of 100 V magnitude occurs in the racetrack coil, the current initially rises sharply overcoming the critical current, and then it decreases following noticeable drops to a stable value (Figure \ref{f.100V_cur}). As we detail in the next paragraph, the initial current increase is accompanied by rapid temperature rise, which results in a transition of a few turns of the coil from the superconducting to the normal state (Figure \ref{f.100V_temp}(a,b)). Afterwards, heat generated in these turns begin diffusing into neighboring turns, leading to observable drops in current (Figure \ref{f.100V_cur}).

\begin{figure}[htp]
\subfloat[initial ramp]{
  {\includegraphics[trim=1.6cm 0 1.4cm 0,clip,height=4.1cm]{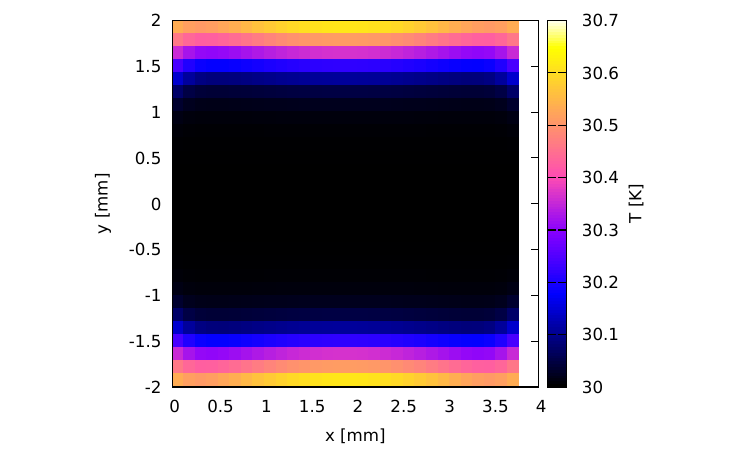}}
}
\subfloat[initial ramp]{
  {\includegraphics[trim=1.6cm 0 1.9cm 0,clip,height=4.1cm]{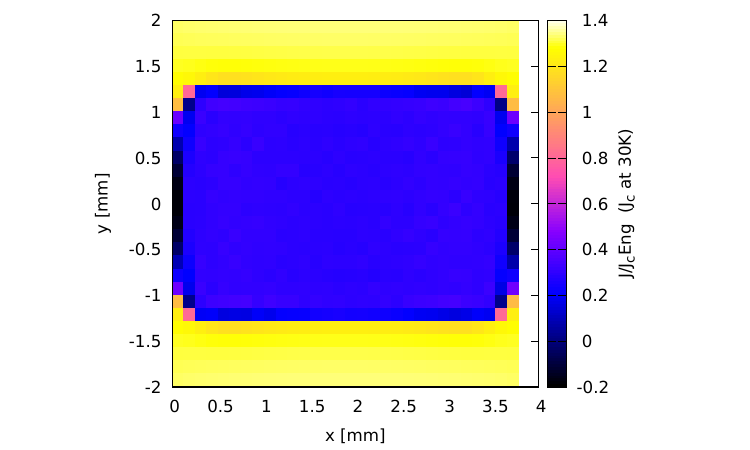}}
}
\subfloat[initial ramp]{
  {\includegraphics[trim=1.6cm 0 1.4cm 0,clip,height=4.1cm]{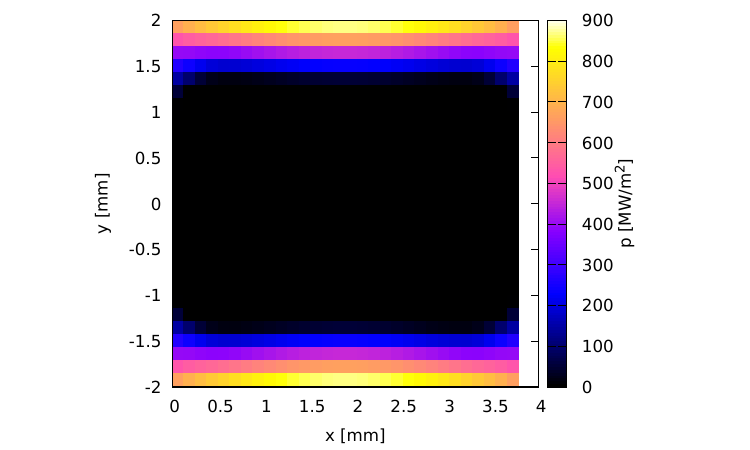}}
}

\hspace{0mm}
\subfloat[current peak]{
  {\includegraphics[trim=1.6cm 0 1.4cm 0,clip,height=4.1cm]{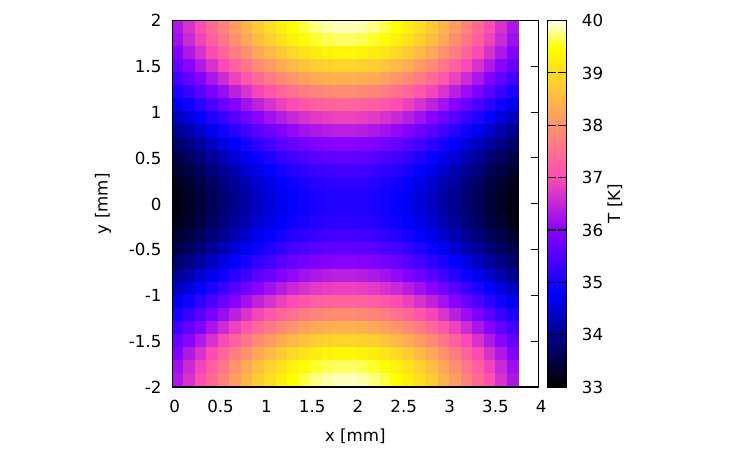}}
}
\subfloat[current peak]{
  {\includegraphics[trim=1.6cm 0 1.75cm 0,clip,height=4.1cm]{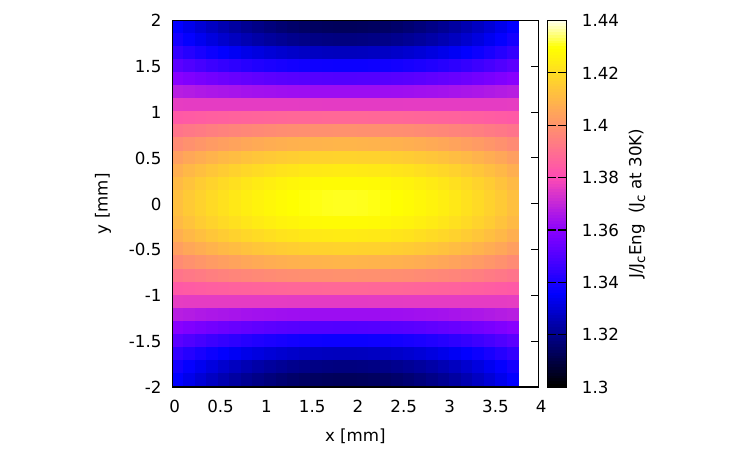}}
}
\subfloat[current peak]{
  {\includegraphics[trim=1.6cm 0 1.4cm 0,clip,height=4.1cm]{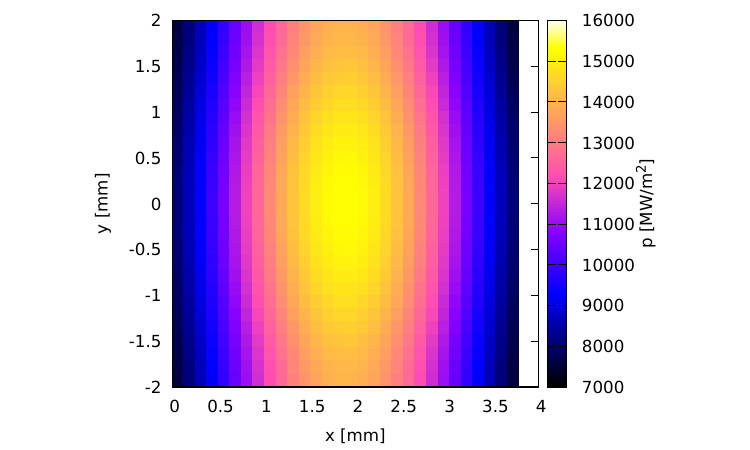}}
}

\hspace{0mm}
\subfloat[first drop]{
  {\includegraphics[trim=1.6cm 0 1.4cm 0,clip,height=4.1cm]{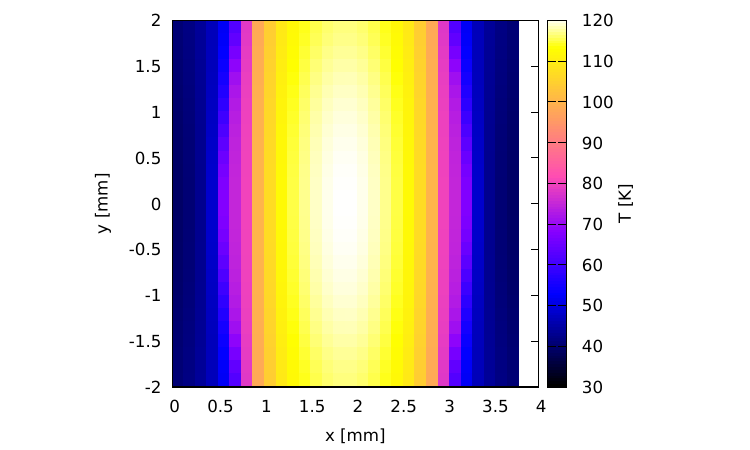}}
}
\subfloat[first drop]{
  {\includegraphics[trim=1.6cm 0 1.9cm 0,clip,height=4.1cm]{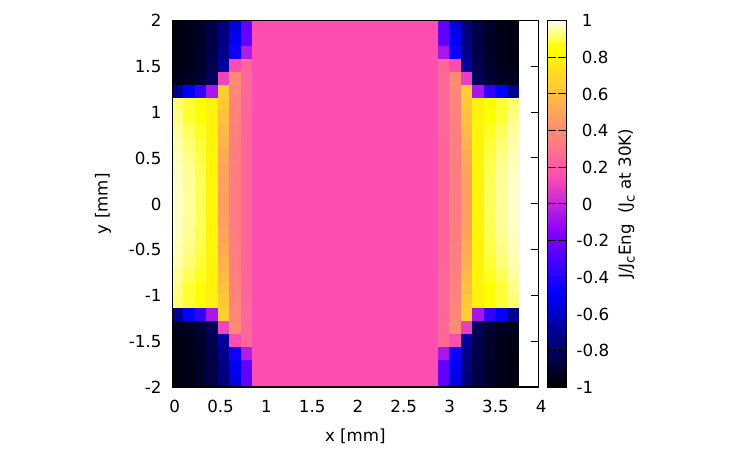}}
}
\subfloat[first drop]{
  {\includegraphics[trim=1.6cm 0 1.4cm 0,clip,height=4.1cm]{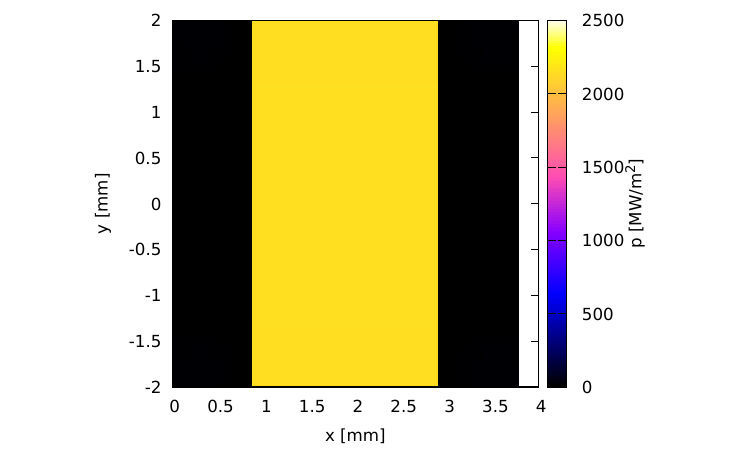}}
}

\hspace{0mm}
\subfloat[third drop]{
  {\includegraphics[trim=1.6cm 0 1.4cm 0,clip,height=4.1cm]{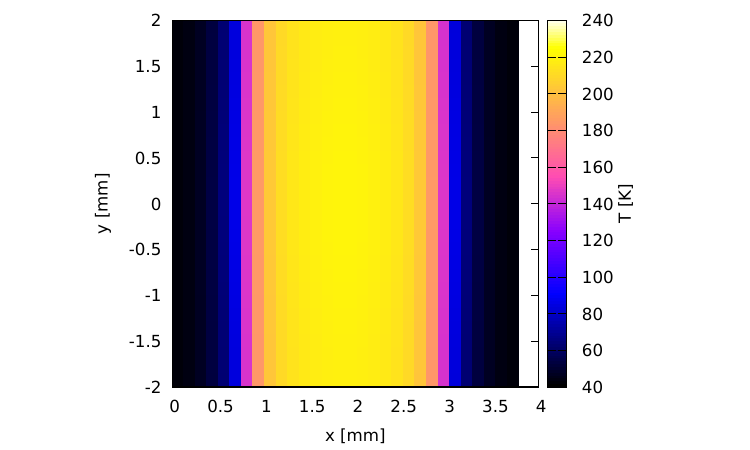}}
}
\subfloat[third drop]{
  {\includegraphics[trim=1.6cm 0 1.9cm 0,clip,height=4.1cm]{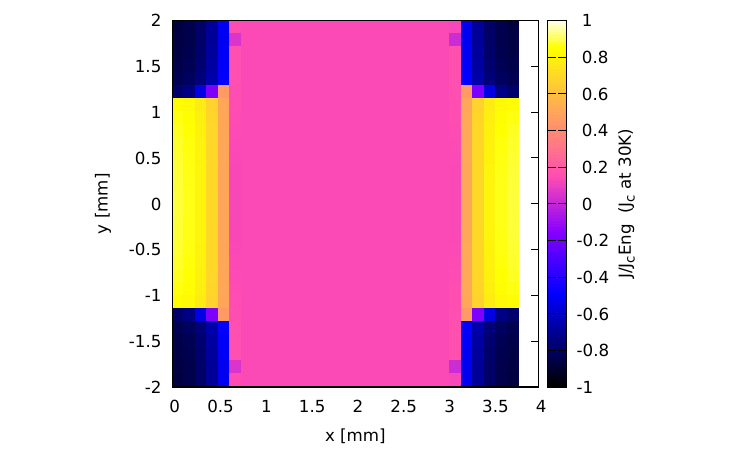}}
}
\subfloat[third drop]{
  {\includegraphics[trim=1.6cm 0 1.4cm 0,clip,height=4.1cm]{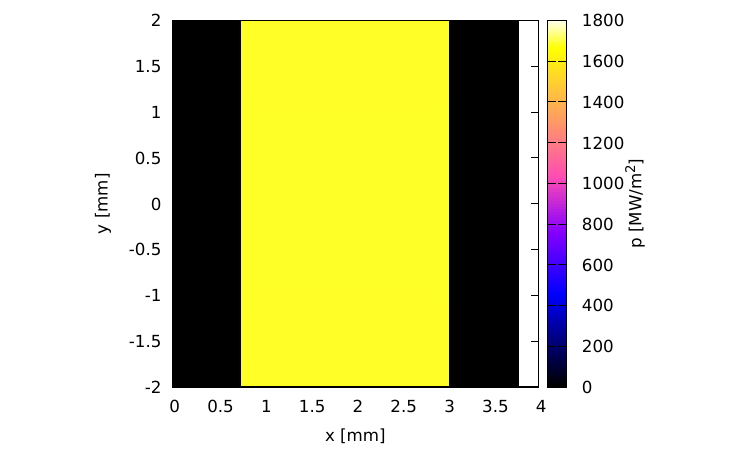}}
}

\caption{Coil cross-section showing maps of temperature (left), current density (middle), and power loss density (right) for adiabatic conditions at 100 V amplitude of DC voltage for different times: (a,b,c) at point A, (d,e,f) at at point B, (g,h,i) at point C, and (j,k,l) at point D in figures \ref{f.100V_cur} and \ref{f.100V_temp}.}
\label{f.maps_silver_100Vdc}
\end{figure}

For the detailed analysis, we study the interaction between current density, temperature rise, and the power loss distributions at different time steps. At the initial ramp (point A in figures \ref{f.100V_cur} and \ref{f.100V_temp}), we see a small temperature rise due to penetrating screening currents and corresponding power loss (Figure \ref{f.maps_silver_100Vdc}(a,b,c)). The higher temperatures observed at the top and bottom of the coil result from AC losses generated during the initial current ramp. These losses, which are due to superconducting screening currents, lead to localized heating in these regions. At the peak of the current (point B in figures \ref{f.100V_cur} and \ref{f.100V_temp}), there are no screening currents because the current density ($J$) exceeds the critical current density ($J_c$) across the entire cross-section, and hence the whole coil is in over-critical state ($I$ above each tape $I_c$). This over-critical state leads to high power dissipation throughout the cross-section (figure \ref{f.maps_silver_100Vdc}(d,e,f)). However, this dissipation is not uniform, being higher at the central $x$ coordinate of the coil. The cause seems to be that for any given $y$, the temperature is higher at the central $x$ than at the edges, which causes a higher ratio of $J$ over the local $J_c$ and consequent higher power loss. Next, we analyze the state after the current drop of point C in figures \ref{f.100V_cur} and \ref{f.100V_temp} (figure \ref{f.maps_silver_100Vdc}(g,h,i)). When the temperature exceeds the critical temperature ($T>T_c$), the material transits to its normal resistive state, causing the net current ($I$) to decrease significantly since the overall resistance increases but the voltage is kept constant. This also causes a decrease in current density in the normal conducting turns. In the superconducting turns, the current decrease causes changing magnetic flux, which induces screening currents. Almost all dissipation occurs in the region where $T>T_c$ because the material exhibits finite resistance, leading to significant Joule heating. The stepwise current drop (point D in figure \ref{f.100V_cur} and \ref{f.100V_temp}) indicates the propagation of a normal zone in the coil. Each step corresponds to a turn (or symmetric pair of turns) losing superconductivity, one on the right and another on the left of the normal zone. This transition increases resistance and Joule heating, further raising the temperature and expanding the normal region (figure \ref{f.maps_silver_100Vdc}(j,k,l)).


\begin{figure}[htp]
\centering
\includegraphics[trim=0 0 0.35cm 0,clip,width=10 cm]{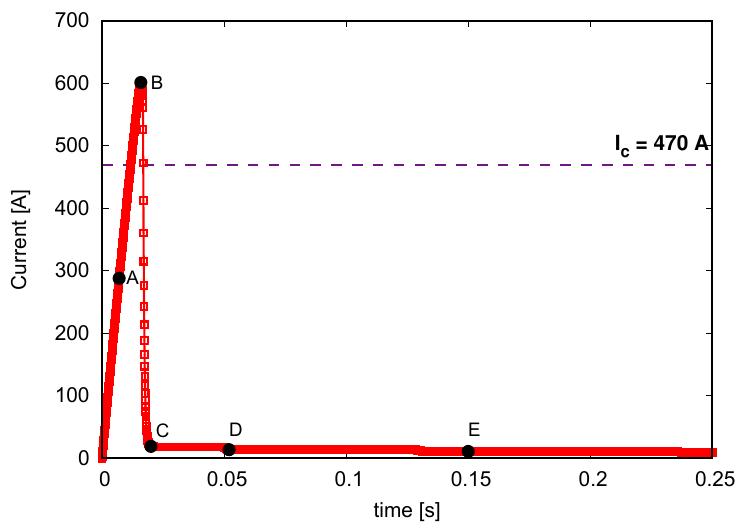}
\caption{Total current in racetrack coil for 10 V DC input with no heat exchange with its surroundings. }
\label{f.10Vdc_cur}
\end{figure}

\begin{figure}[htp]
\centering
\includegraphics[trim=0 0 0 0,clip,width=10 cm]{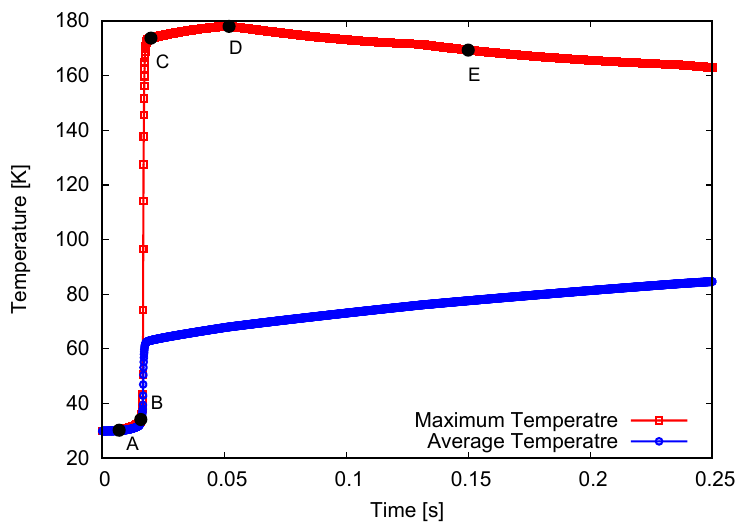}
\caption{Temperature rise in racetrack coil for 10 V DC input with no heat exchange with the cryogenic liquid. }
\label{f.10Vdc_temp}
\end{figure}

We further reduce the magnitude of short-circuit DC voltage to 10 V. The superconducting racetrack coil exhibits a characteristic response pattern consistent with previously observed behaviors at higher fault voltages. The current initially shows the same trend as for higher short-circuit DC voltages (figure \ref{f.10Vdc_cur}): a ramp increase followed by a steep drop. However, after the first significant drop in current to around 19.5 A, a peak in temperature appears (see point D in figure \ref{f.10Vdc_cur} and \ref{f.10Vdc_temp}). After the current peak, there appear small current drops (figure \ref{f.10Vdc_cur}), as for 100 V.

\begin{figure}[htp]
\subfloat[initial ramp]{
  {\includegraphics[trim=1.6cm 0.4cm 1.4cm 0,clip,height=3.5cm]{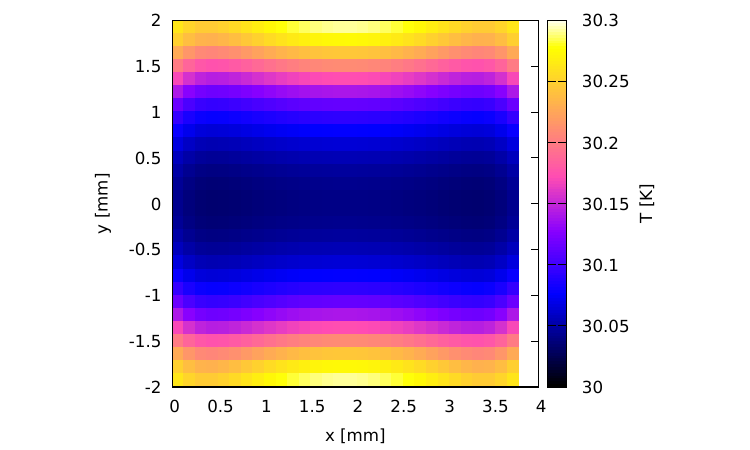}}
}
\subfloat[initial ramp]{
  {\includegraphics[trim=1.6cm 0.4cm 1.9cm 0,clip,height=3.5cm]{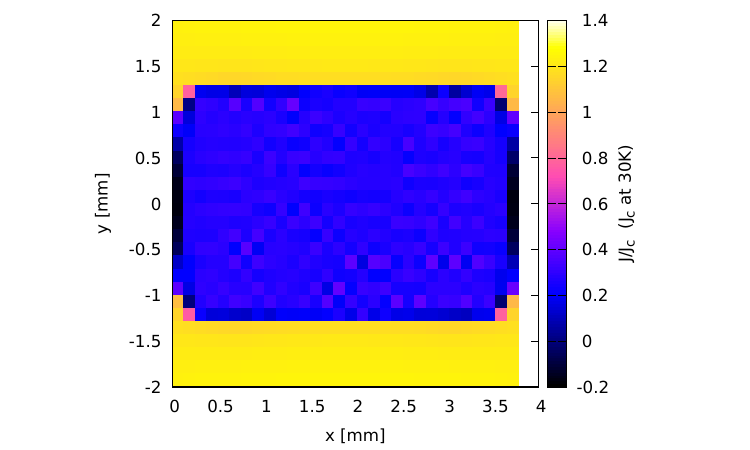}}
}
\subfloat[initial ramp]{
  {\includegraphics[trim=1.6cm 0.4cm 1.4cm 0,clip,height=3.5cm]{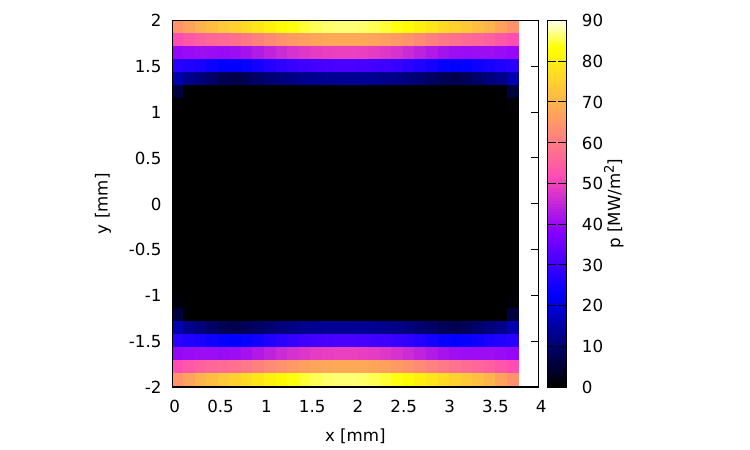}}
}

\hspace{0mm}
\subfloat[current peak]{
  {\includegraphics[trim=1.6cm 0.4cm 1.6cm 0,clip,height=3.5cm]{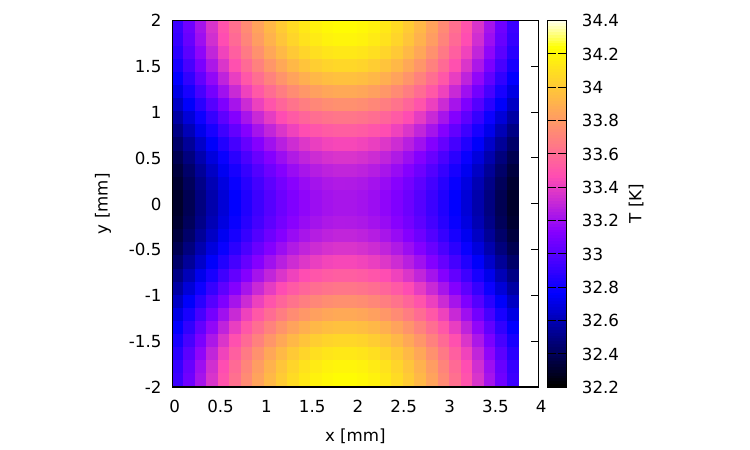}}
}
\subfloat[current peak]{
  {\includegraphics[trim=1.6cm 0.4cm 1.6cm 0,clip,height=3.5cm]{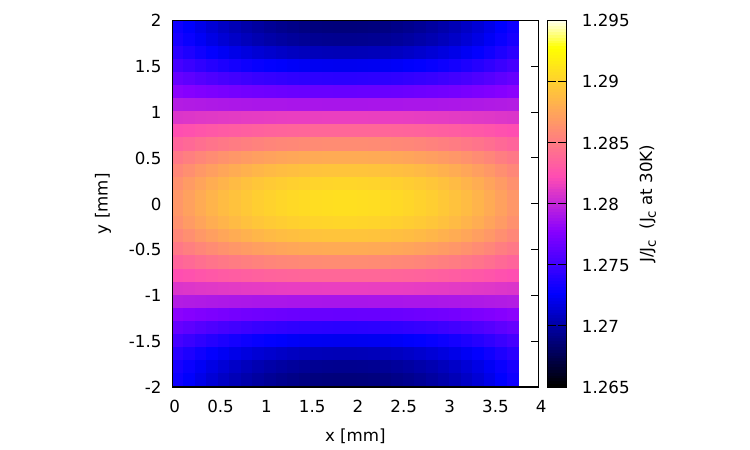}}
}
\subfloat[current peak]{
  {\includegraphics[trim=1.6cm 0.4cm 1.6cm 0,clip,height=3.5cm]{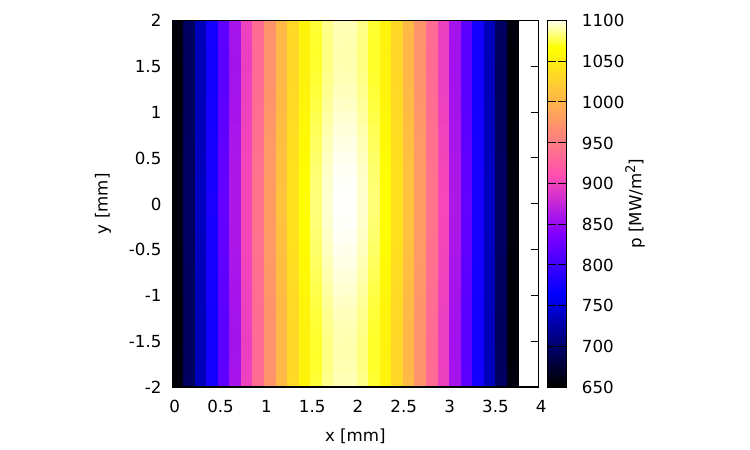}}
}

\hspace{0mm}
\subfloat[first drop]{
  {\includegraphics[trim=1.6cm 0.4cm 1.4cm 0,clip,height=3.5cm]{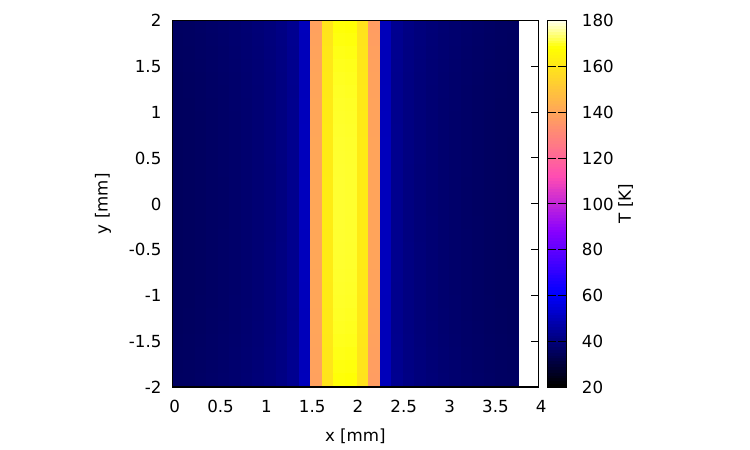}}
}
\subfloat[first drop]{
  {\includegraphics[trim=1.6cm 0.4cm 1.9cm 0,clip,height=3.5cm]{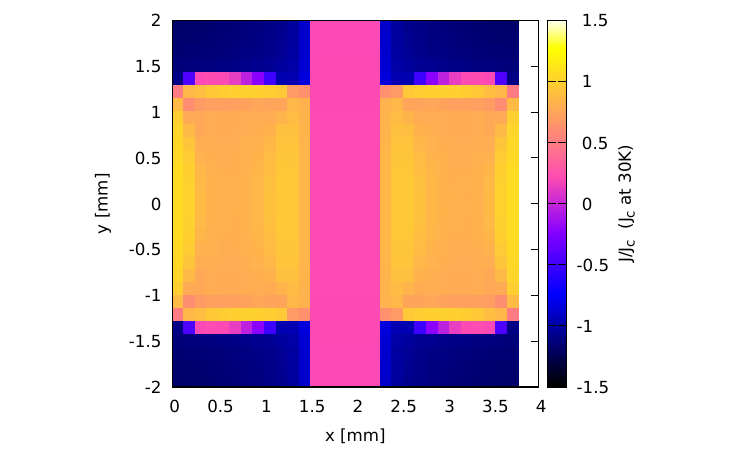}}
}
\subfloat[first drop]{
  {\includegraphics[trim=1.6cm 0.4cm 1.4cm 0,clip,height=3.5cm]{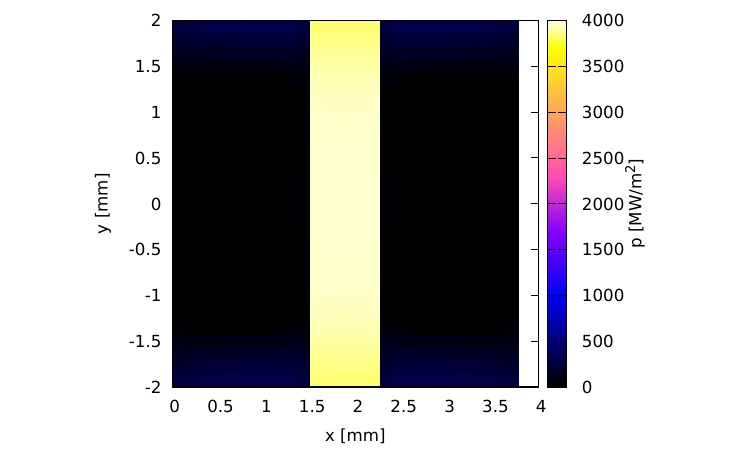}}
}

\hspace{0mm}
\subfloat[second current peak]{
  {\includegraphics[trim=1.6cm 0.4cm 1.4cm 0,clip,height=3.5cm]{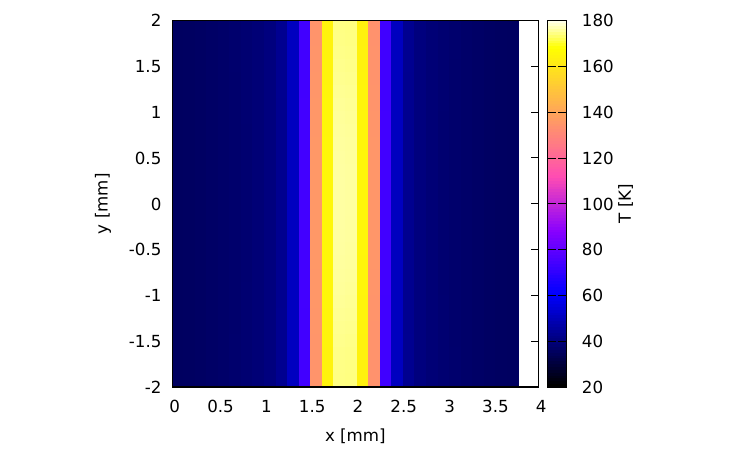}}
}
\subfloat[second current peak]{
  {\includegraphics[trim=1.6cm 0.4cm 1.9cm 0,clip,height=3.5cm]{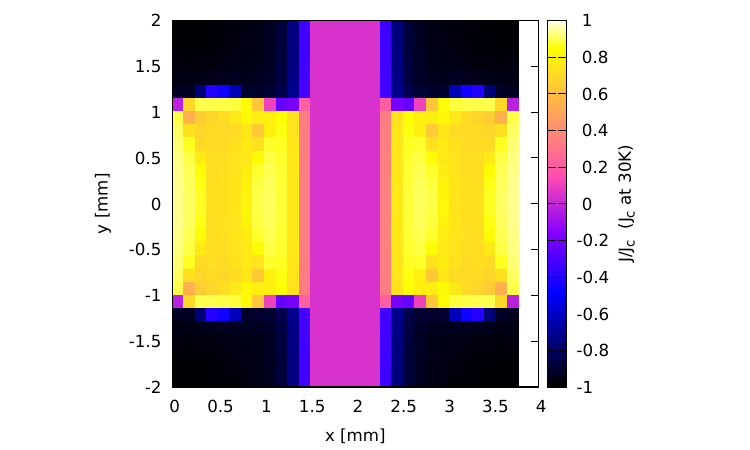}}
}
\subfloat[second current peak]{
  {\includegraphics[trim=1.6cm 0.4cm 1.4cm 0,clip,height=3.5cm]{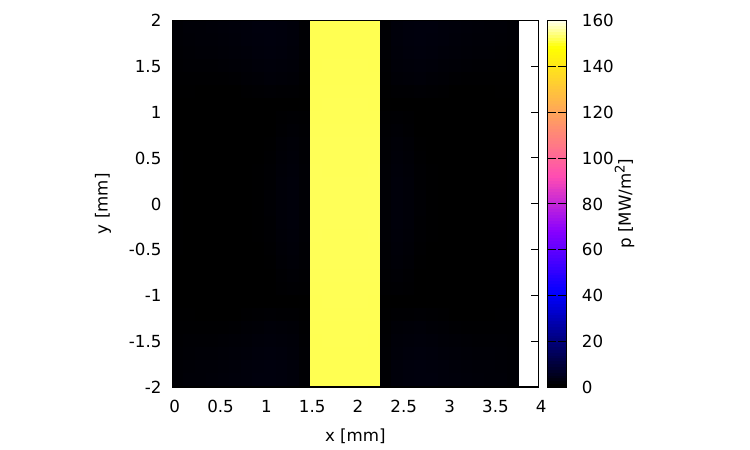}}
}

\hspace{0mm}
\subfloat[stable current]{
  {\includegraphics[trim=1.6cm 0.4cm 1.4cm 0,clip,height=3.5cm]{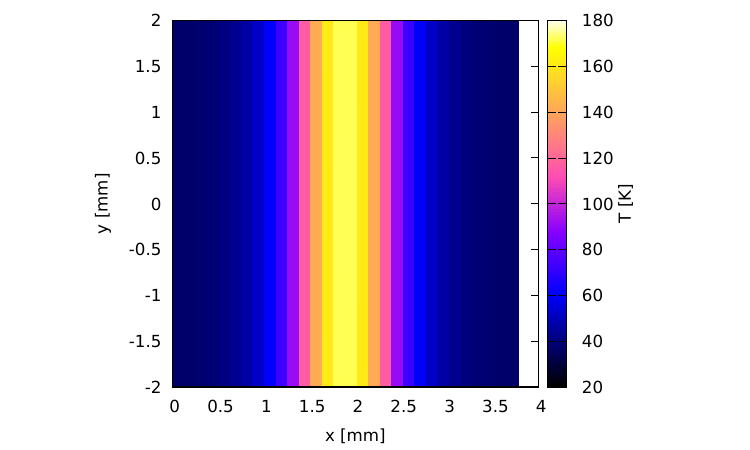}}
}
\subfloat[stable current]{
  {\includegraphics[trim=1.6cm 0.4cm 1.6cm 0,clip,height=3.5cm]{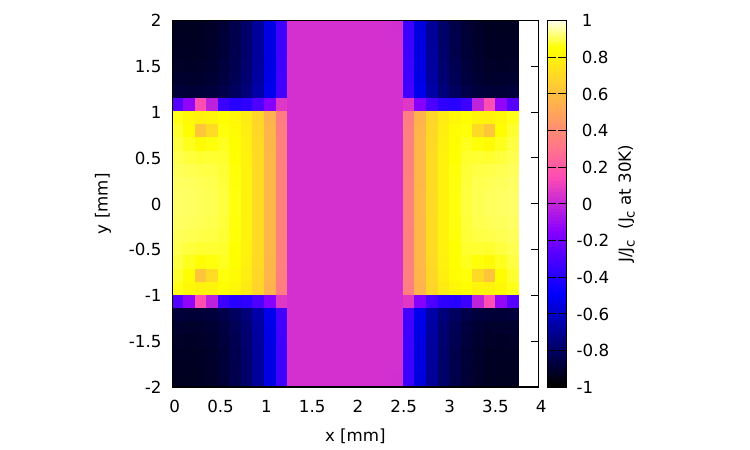}}
}
\subfloat[stable current]{
  {\includegraphics[trim=1.6cm 0.4cm 1.4cm 0,clip,height=3.5cm]{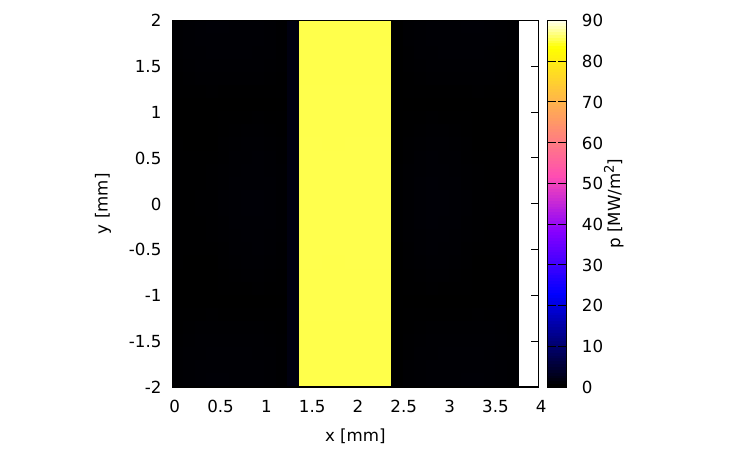}}
}
\caption{Coil cross-section showing maps of temperature (left), current density (middle), and power loss density (right) for adiabatic conditions at 10 V amplitude of DC voltage for different times: (a,b,c) at point A, (d,e,f) at point B, (g,h,i) at point C, (j,k,l) at point D, and (m,n,o) at point E in figures \ref{f.10Vdc_cur} and \ref{f.10Vdc_temp}.}
\label{f.maps_silver_10Vdc}
\end{figure}

Next we analyze the temperature, normalized current density, and power loss distributions at the key time steps (figure \ref{f.maps_silver_10Vdc}). During the initial ramp (point A in figures \ref{f.10Vdc_cur} and \ref{f.10Vdc_temp}), the behavior qualitatively remains the same as the previous cases: we see a small temperature rise due to screening currents, and corresponding power loss (Figure \ref{f.maps_silver_10Vdc}(a,b,c)). At the current peak (point B in figures \ref{f.10Vdc_cur} and \ref{f.10Vdc_temp}), there is temperature rise in the whole coil because at this time the current is above the critical current of any turn, which causes high power loss (Figure \ref{f.maps_silver_10Vdc}(d,e,f)). Furthermore, just after the first drop of current, the temperature is above the critical temperature at the central turns ($T > T{_c}$), and hence these turns become normal (Figure \ref{f.maps_silver_10Vdc}(g,h,i)). This causes a strong suppression of the coil net current, as well as the current density at the central turns. Afterwards, the maximum temperature decreases with a roughly exponential decay (region between C and D in figure \ref{f.10Vdc_cur}). The reasons are: first, dissipation is low due to low transport current; second, the heat diffuses from the central turns, where the temperature is the highest, to their neighbors (compare figures \ref{f.maps_silver_10Vdc}(j,k,l) with figures \ref{f.maps_silver_10Vdc}(m,n,o)). After the initial drop in current (point C in figure \ref{f.10Vdc_cur}), there appear smaller drops in current due to propagation of the normal zone to neighboring turns. Each drop in current corresponds to a pair of turns that became normal (compare figures \ref{f.maps_silver_10Vdc}(g,h,i) with \ref{f.maps_silver_10Vdc}(j,kl) and \ref{f.maps_silver_10Vdc}(m,n,o)).


\begin{figure}[htp]
\centering
\includegraphics[trim=0 0 0 0,clip,width=10 cm]{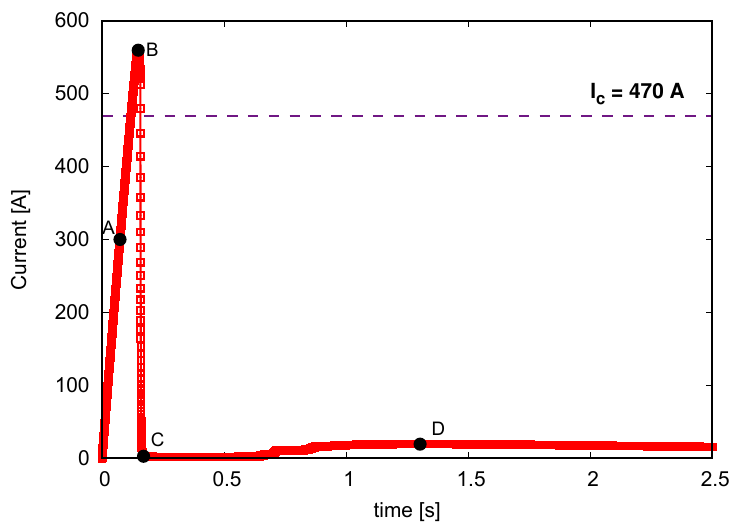}\\
\caption{Total current in racetrack coil for 1 V DC input with no heat exchange with its surroundings.}
\label{f.1V_cur1}
\end{figure}

\begin{figure}[htp]
\centering
  \includegraphics[trim=0 0 0 0,clip,width=10 cm]{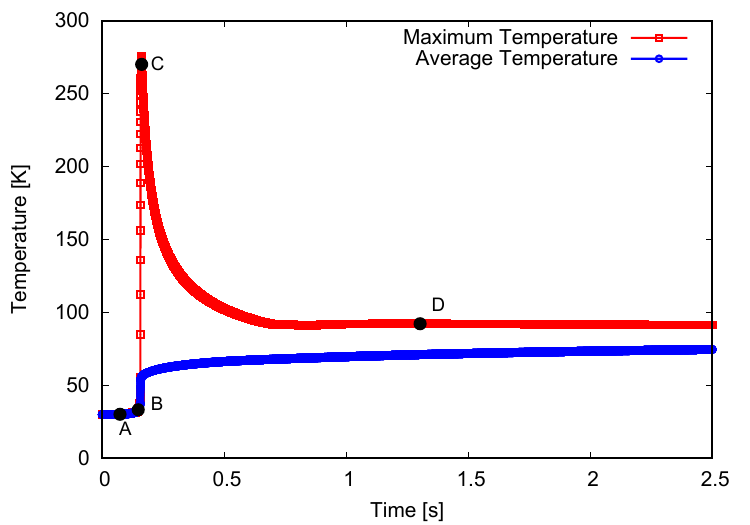} \\
\caption{Temperature rise in racetrack coil for 1 V DC input with no heat exchange with the cryogenic liquid. } 
\label{f.1V_temp1}
\end{figure}

When a 1\,V in DC occurs in the racetrack coil, the current initially ramps up sharply exceeding the critical current and then drops sharply, similar to higher voltage cases (figure \ref{f.1V_cur1}). As the current approaches the critical threshold, parts of the coil transits to a resistive state due to localized higher current density. This leads to a sharp current drop and a rapid temperature rise caused by Joule heating in the normal zones (figure \ref{f.1V_temp1}). Cooling of the hottest turns occurs due to internal thermal diffusion, and hence the maximum temperature in the racetrack coil decreases as the current remains near to zero.  This cooling, results in a reduction of the number of turns in normal state, causing a small increase in current and a peak (point D in figure \ref{f.1V_cur1}).

\begin{figure}[htp]
\subfloat[initial ramp]{
  {\includegraphics[trim=1.6cm 0 1.4cm 0,clip,height=4.1cm]{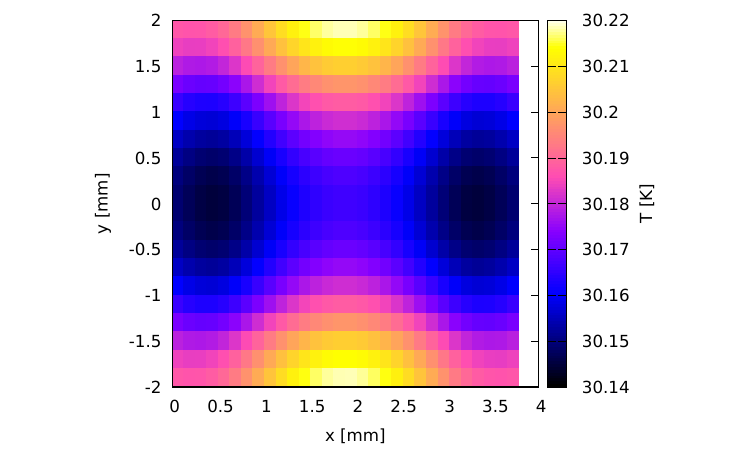}}
}
\subfloat[initial ramp]{
  {\includegraphics[trim=1.6cm 0 1.9cm 0,clip,height=4.1cm]{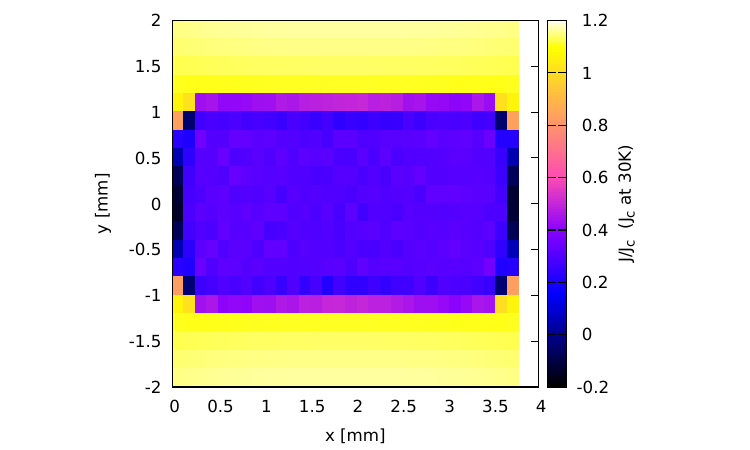}}
}
\subfloat[initial ramp]{
  {\includegraphics[trim=1.6cm 0 1.4cm 0,clip,height=4.1cm]{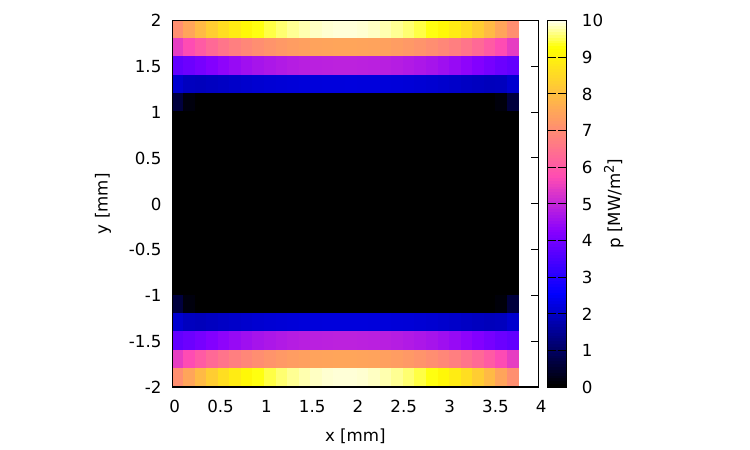}}
}

\hspace{0mm}
\subfloat[first current peak]{
  {\includegraphics[trim=1.6cm 0 1.4cm 0,clip,height=4.1cm]{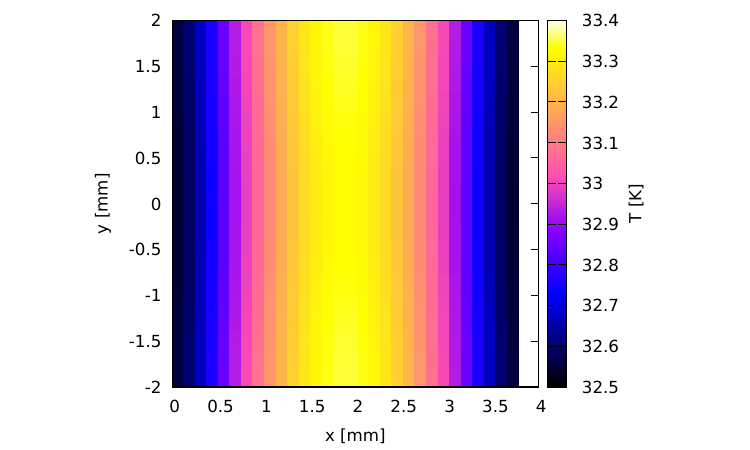}}
}
\subfloat[first current peak]{
  {\includegraphics[trim=1.6cm 0 1.6cm 0,clip,height=4.1cm]{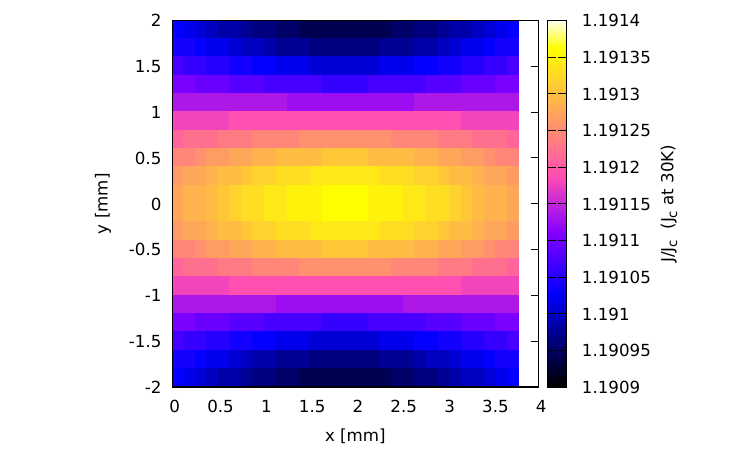}}
}
\subfloat[first current peak]{
  {\includegraphics[trim=1.6cm 0 1.4cm 0,clip,height=4.1cm]{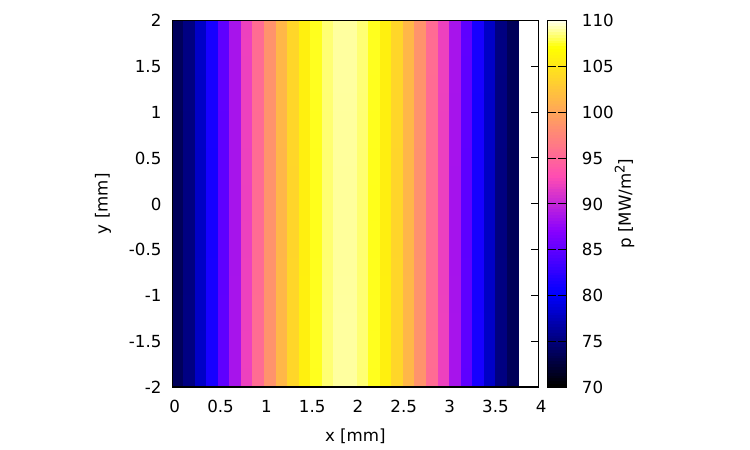}}
}

\hspace{0mm}
\subfloat[first current drop]{
  {\includegraphics[trim=1.6cm 0 1.4cm 0,clip,height=4.1cm]{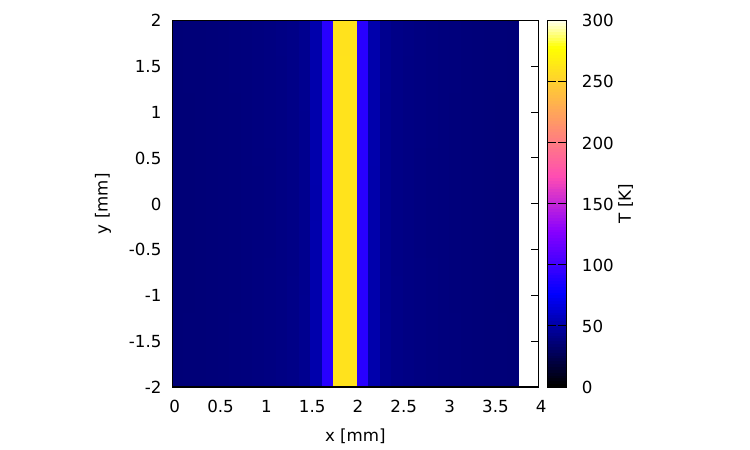}}
}
\subfloat[first current drop]{
  {\includegraphics[trim=1.6cm 0 1.9cm 0,clip,height=4.1cm]{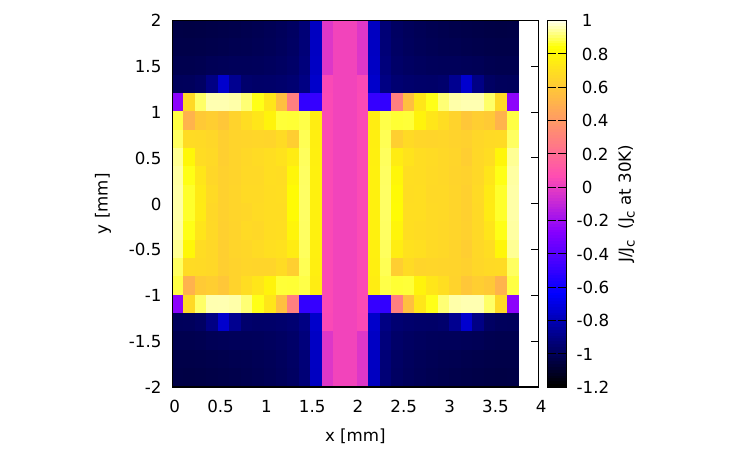}}
}
\subfloat[first current drop]{
  {\includegraphics[trim=1.6cm 0 1.4cm 0,clip,height=4.1cm]{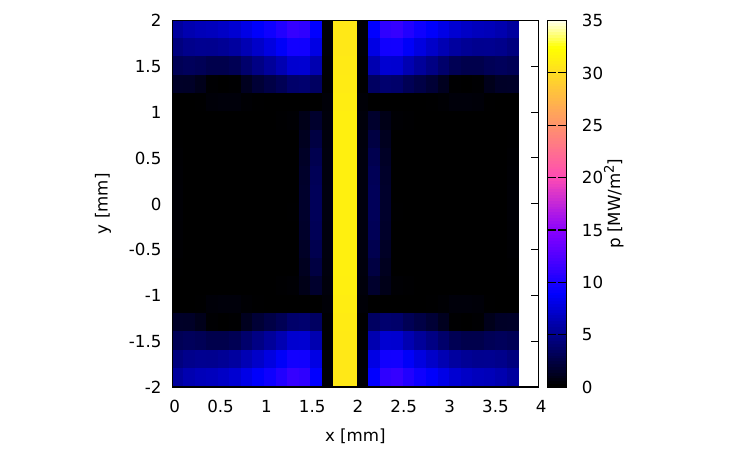}}
}

\hspace{0mm}
\subfloat[second current peak]{
  {\includegraphics[trim=1.6cm 0 1.4cm 0,clip,height=4.1cm]{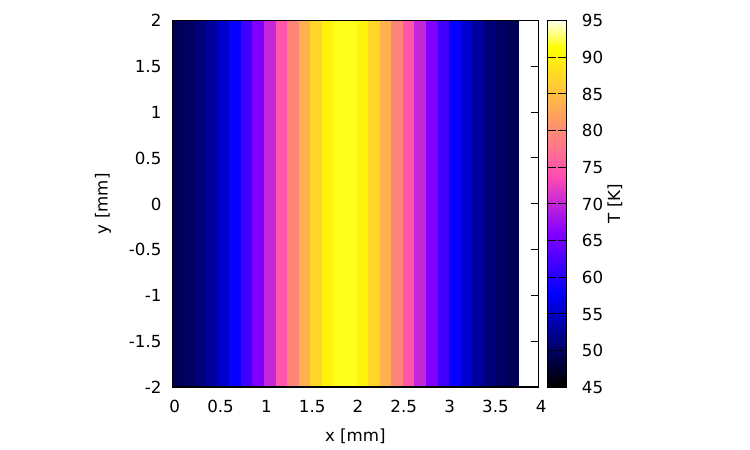}}
}
\subfloat[second current peak]{
  {\includegraphics[trim=1.6cm 0 1.9cm 0,clip,height=4.1cm]{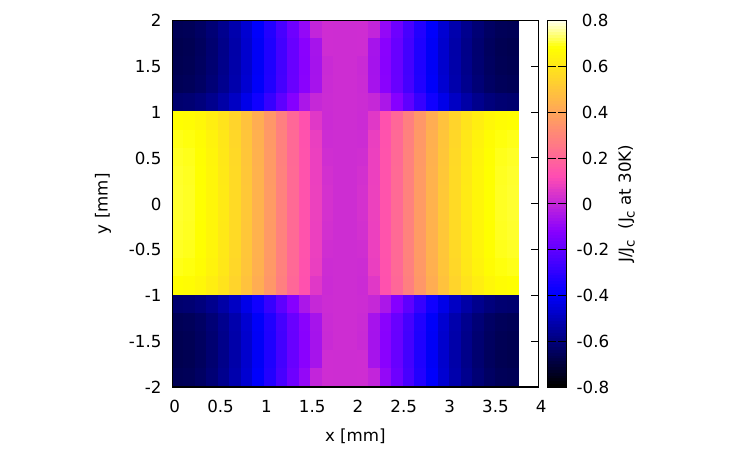}}
}
\subfloat[second current peak]{
  {\includegraphics[trim=1.6cm 0 1.4cm 0,clip,height=4.1cm]{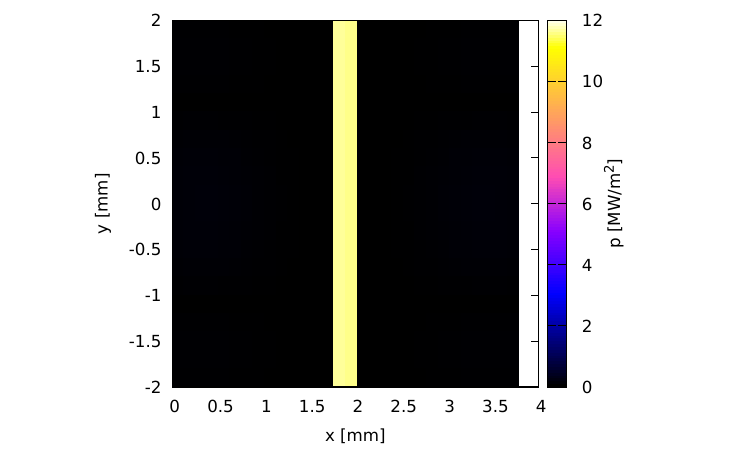}}
}

\caption{Coil cross-section showing maps of temperature (left), current density (middle), and power loss density (right) for adiabatic conditions at 1 V amplitude of DC voltage for different times: (a,b,c) at point A, (d,e,f) at point B, (g,h,i) at point C, and (j,k,l) at point D in figures \ref{f.1V_cur1} and \ref{f.1V_temp1}.}
\label{f.maps_silver_1Vdc1}
\end{figure}

Next we analyze the temperature, current density, and power loss distributions at specific time steps (figure \ref{f.maps_silver_1Vdc1}). At the initial current ramp (point A in figure \ref{f.1V_cur1} and \ref{f.1V_temp1}), the behavior remains qualitatively the same as previous cases. A small temperature rise occurs due to screening currents and corresponding low power loss in the cross-section of the racetrack coil (figure \ref{f.maps_silver_1Vdc1}(a,b,c)). We further analyze the behavior at the current peak (point B in figure \ref{f.1V_cur1} and \ref{f.1V_temp1}). At this time, the cross section of the racetrack coil starts warming up due to the fact that current exceeds the critical current value ($J > J_c$) (figure \ref{f.maps_silver_1Vdc1}(d,e,f)). At the center of the coil, the temperature is higher because: first, AC loss at the initial ramp was higher at the center (figure 15c); and second, zones at higher temperature experience lower $J_c$, which causes higher local power loss (figure 15f). As a consequence, there is a positive feedback loop at the central turns that results in a sharp temperature rise there. Then, the central turns become normal resistive, which causes the first current drop (point C in figure 13 and 14). At this point, the central turns present $T>T_c$, $J\ll J_c$ and higher Joule loss than the other turns (figure 15(g,h,i)). Afterwards, the heat diffuses to the neighboring turns, which causes that the two central turns are slightly below $T_c=92$ K; and hence their critical current becomes non-zero, although still small (figure \ref{f.maps_silver_1Vdc1}(j,k,l)). This causes the second, shallow, peak in current of figure \ref{f.1V_cur1}. 


\subsection{Conduction Cooling}

\begin{figure}[htp]
\centering
\includegraphics[trim=0 0 0 0,clip,width=7 cm]{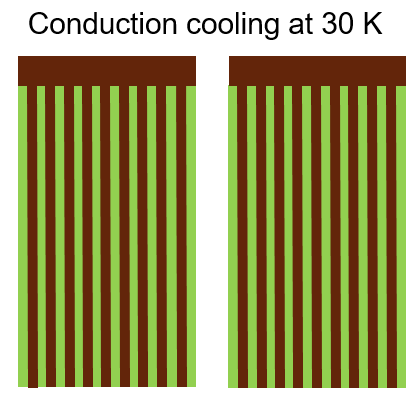}
\caption{ Sketch of the cross section of the racetrack coil with cooling from the top, where green color represents superconductor with all metallic layers, and red color represents polyimide (kapton) insulation layer.}
\label{f.topcooling}
\end{figure}

In this study, we explore how conduction cooling from the top surface of the racetrack coil improves its performance. To achieve this, a thin 10-micron layer of polyimide (Kapton) is placed on the top of the coil, with the temperature on its surface carefully maintained at 30 K throughout the process (Figure \ref{f.topcooling}). This simulates conduction cooling from one side.

\begin{figure}[htp]
\centering
\includegraphics[trim=0 0 0 0,clip,width=9 cm]{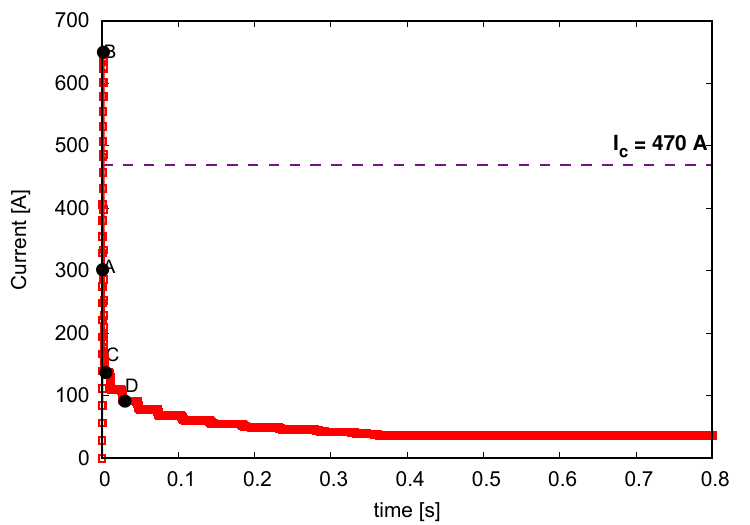}\\
\includegraphics[trim=0 0 0 0,clip,width=9 cm]{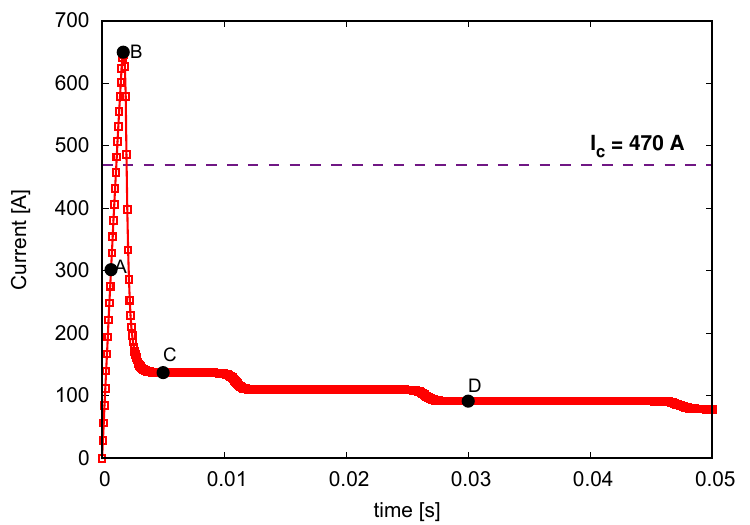}
\caption{ (a) Total current in racetrack coil for 100 V DC input with with cooling from the top at $30$ K. (b) Zoomed sketch of total current from $0$ to $50$ ms. }
\label{f.100V_cur_cooling}
\end{figure}

\begin{figure}[htp]
\centering
\includegraphics[trim=0 0 0 0,clip,width=9 cm]{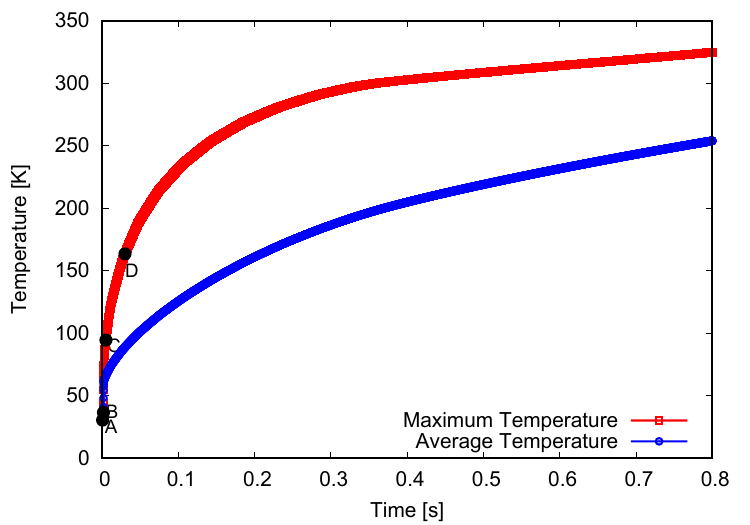}\\
\includegraphics[trim=0 0 0 0,clip,width=9 cm]{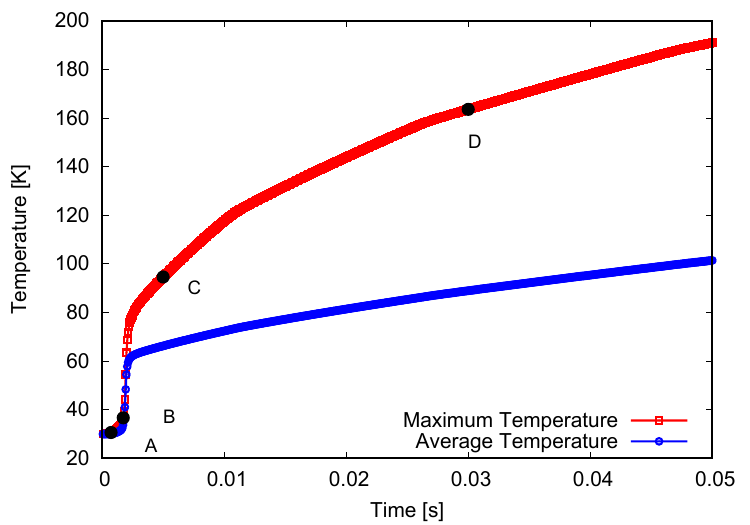}
\caption{ (a) Temperature rise in racetrack coil for 100 V DC input with cooling from the top at $30$ K. (b) Zoomed sketch of temperature from $0$ to $50$ ms.}
\label{f.100V_temp_cooling}
\end{figure}

At 100 V, rapid temperature rise is observed after the initial current ramp (figure \ref{f.100V_cur_cooling} and \ref{f.100V_temp_cooling}), which is due to high Joule heating ($p={\bf J}\cdot{\bf E}$) caused by net current overcoming the critical current at part of the coil. However, the fixed cooling boundary condition (figure \ref{f.topcooling}) effectively extracts the generated heat, resulting in temperature stabilization after approximately $0.8$ s (figure \ref{f.100V_temp_cooling}). This stabilization indicates that the top-surface cooling counteracts the heat generated in the coil, preventing permanent damage due to overheating, which could occur above 400 K. 

\begin{figure}[htp]
\subfloat[initial ramp]{
  {\includegraphics[trim=1.6cm 0 1.4cm 0,clip,height=4.1cm]{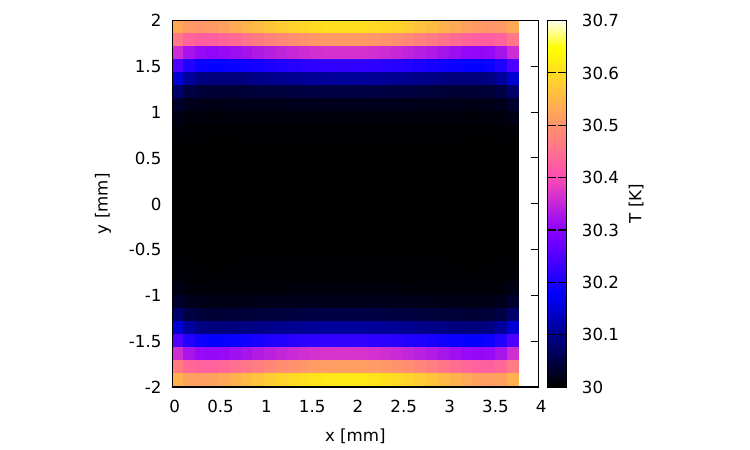}}
}
\subfloat[initial ramp]{
  {\includegraphics[trim=1.6cm 0 1.6cm 0,clip,height=4.1cm]{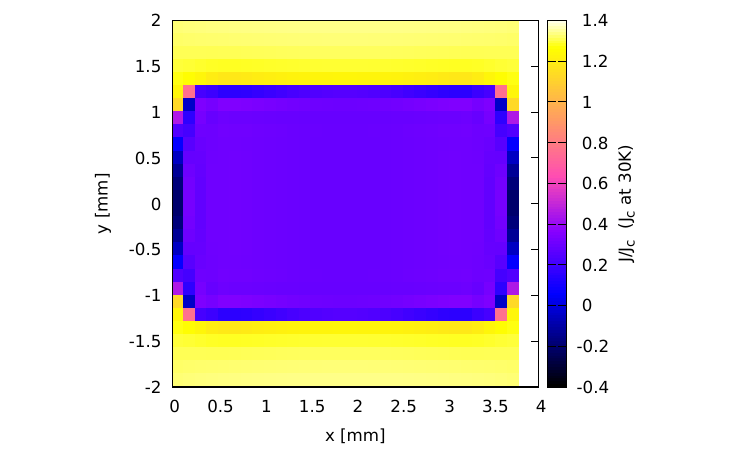}}
}
\subfloat[initial ramp]{
  {\includegraphics[trim=1.6cm 0 1.4cm 0,clip,height=4.1cm]{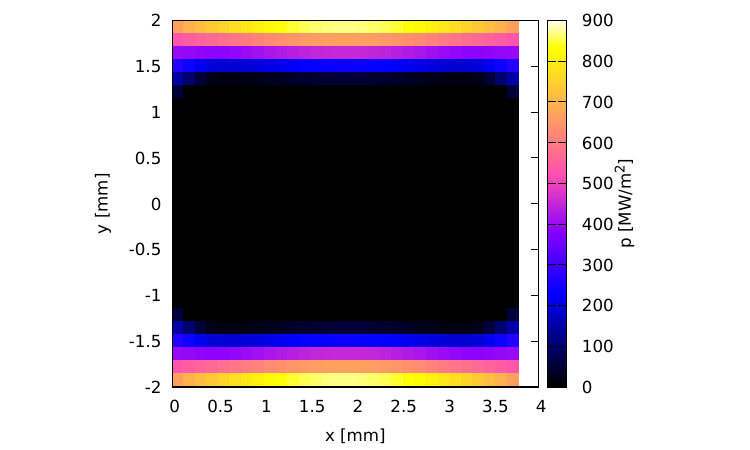}}
}

\hspace{0mm}
\subfloat[current peak]{
  {\includegraphics[trim=1.6cm 0 1.4cm 0,clip,height=4.1cm]{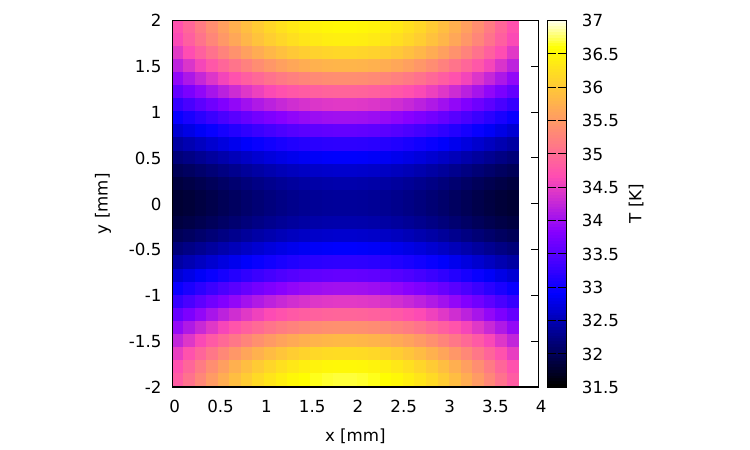}}
}
\subfloat[current peak]{
  {\includegraphics[trim=1.6cm 0 1.6cm 0,clip,height=4.1cm]{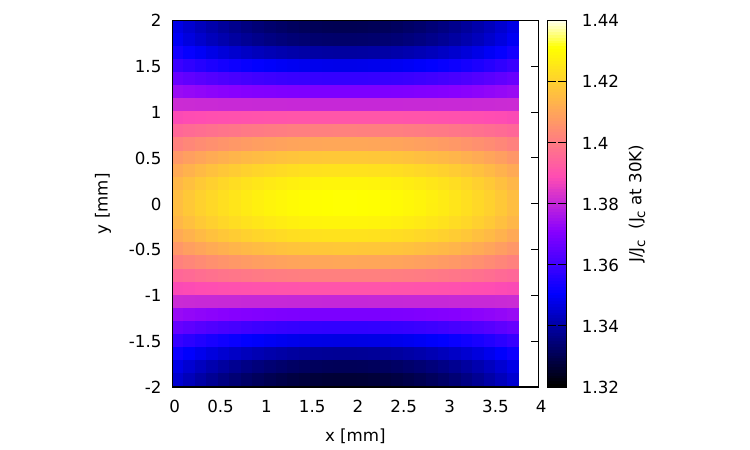}}
}
\subfloat[current peak]{
  {\includegraphics[trim=1.6cm 0 1.4cm 0,clip,height=4.1cm]{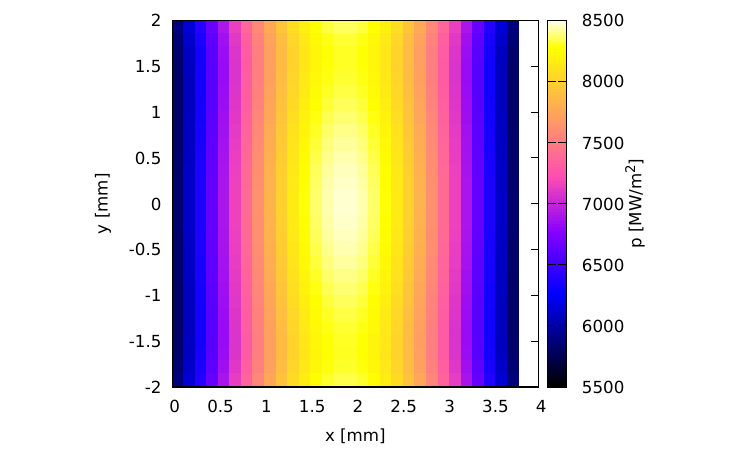}}
}

\hspace{0mm}
\subfloat[first current drop]{
  {\includegraphics[trim=1.6cm 0 1.4cm 0,clip,height=4.1cm]{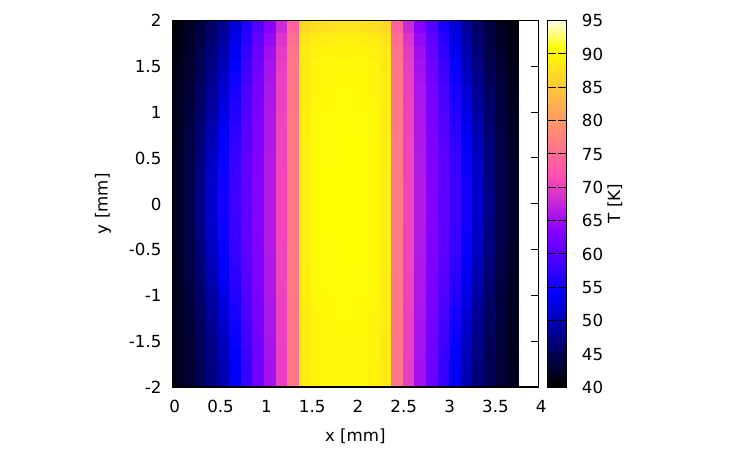}}
}
\subfloat[first current drop]{
  {\includegraphics[trim=1.6cm 0 1.6cm 0,clip,height=4.1cm]{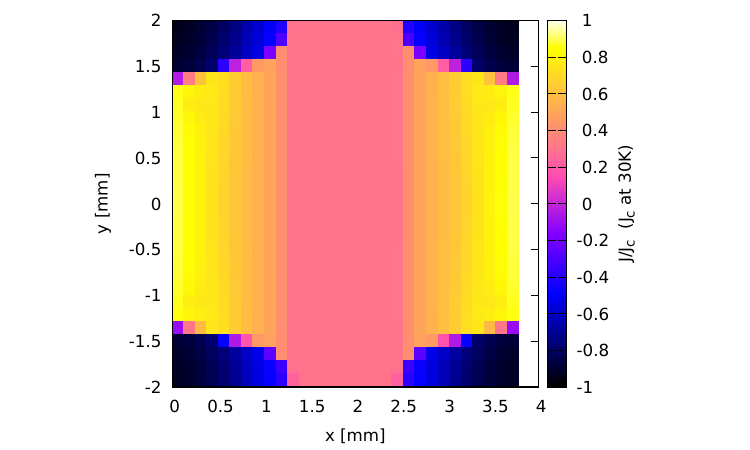}}
}
\subfloat[first current drop]{
  {\includegraphics[trim=1.6cm 0 1.4cm 0,clip,height=4.1cm]{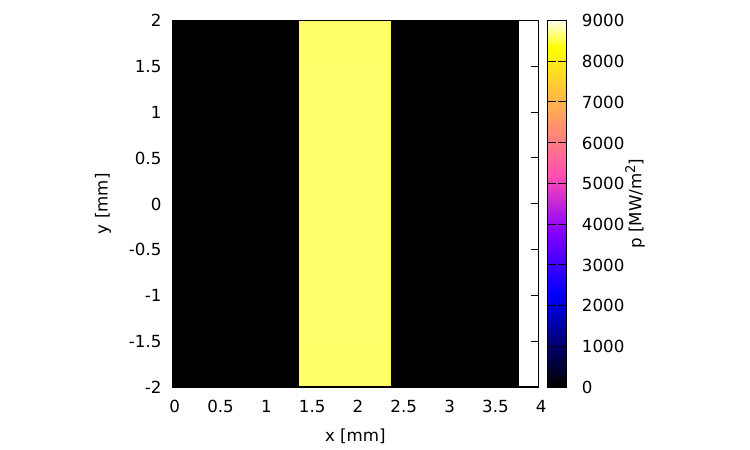}}
}

\hspace{0mm}
\subfloat[third current drop]{
  {\includegraphics[trim=1.6cm 0 1.4cm 0,clip,height=4.1cm]{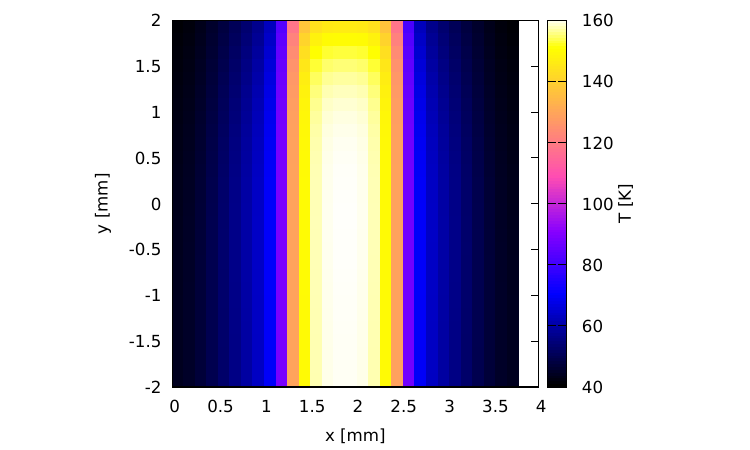}}
}
\subfloat[third current drop]{
  {\includegraphics[trim=1.6cm 0 1.6cm 0,clip,height=4.1cm]{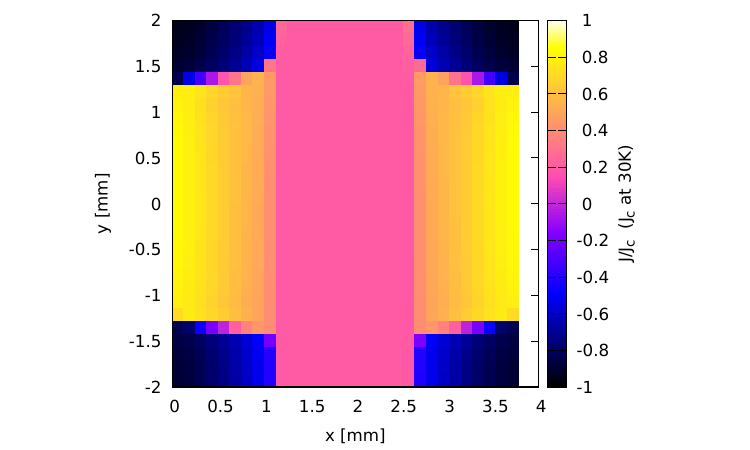}}
}
\subfloat[third current drop]{
  {\includegraphics[trim=1.6cm 0 1.4cm 0,clip,height=4.1cm]{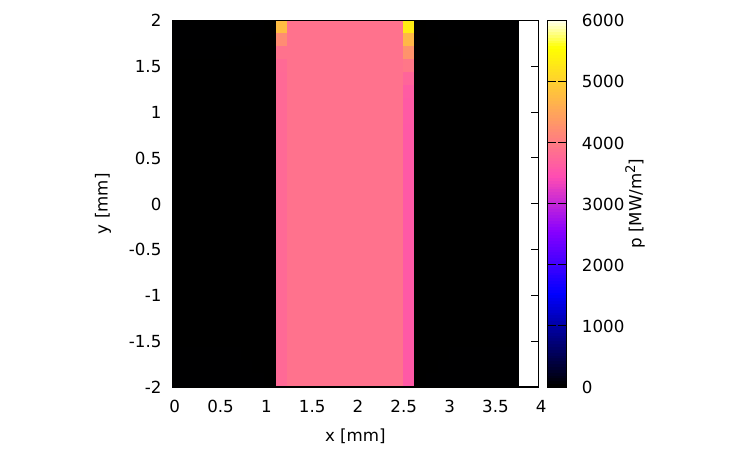}}
}

\caption{Coil cross-section showing maps of temperature (left), current density (middle), and power loss density (right) for top cooling at 100 V amplitude of DC voltage for different times: (a,b,c) at point A, (d,e,f) at point B, (g,h,i) at point C, and (j,k,l) at point D in figures \ref{f.100V_cur_cooling} and \ref{f.100V_temp_cooling}.}
\label{f.maps_100Vdc_cooling}
\end{figure}

Next, we detail the temperature, current, and power distributions. At the initial ramp and up to the current peak the temperature, current density, and power density is qualitatively the same as without cooling (figure \ref{f.maps_100Vdc_cooling}(a,b,c,d,e,f)). The effect of cooling is still not appreciable because of the short elapsed time. At the first current drop, temperature in the central turns is above $T_c$ except the top surface of the coil where cooling is imposed, which experiences temperatures below $T_c$. Over the time, the impact of cooling from the top becomes apparent, as the temperature profile stabilizes under the influence of the fixed cooling boundary (see temperature profiles in figure \ref{f.maps_100Vdc_cooling}(a,d,g,j)). The cooling mechanism avoids thermal runaway, which keeps the maximum temperature well below 400 K, and hence preventing thermal damage. However, the temperature at the central turns is still well above $T_c$. As for the case of no cooling, the current presents several drops caused by the expansion of the normal zone at the center of the coil.


\begin{figure}[htp]
\centering
\includegraphics[trim=0 0 0 0,clip,width=9 cm]{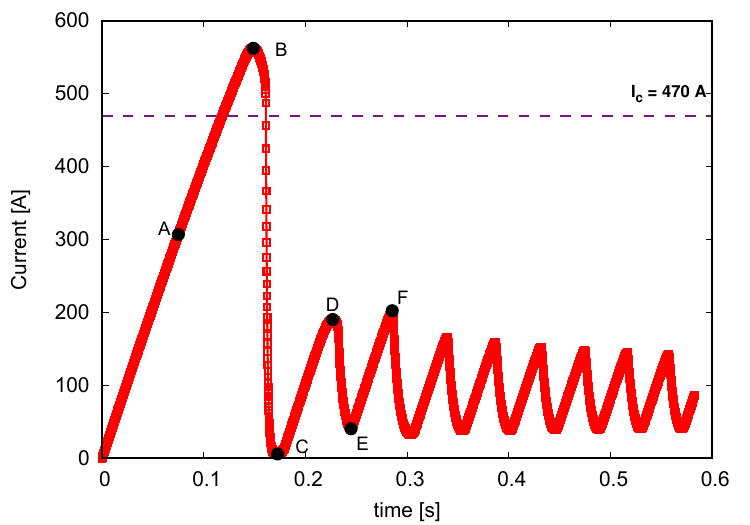}
\caption{Total current in racetrack coil for 1 V DC input with cooling from the top at $30$ K.}
\label{f.1V_cur_cooling}
\end{figure}

\begin{figure}[htp]
\centering
\includegraphics[trim=0 0 0 0,clip,width=9 cm]{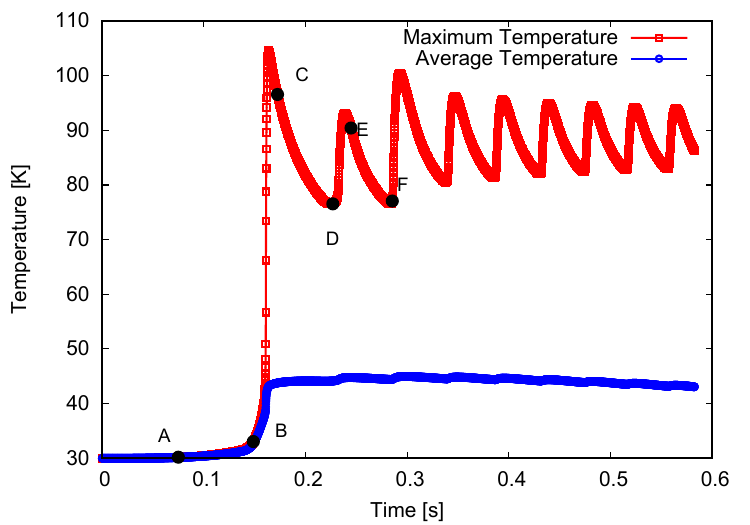}
\caption{Temperature rise in racetrack coil for 1 V DC input with cooling from the top at $30$ K.}
\label{f.1V_temp_cooling}
\end{figure}

Now we study the electromagnetic and thermal response under a 1\,V DC short-circuit condition, with conduction cooling applied only from the top surface at 30\,K. At the initial current ramp (point A), the current increases gradually due to the low voltage magnitude and the inductance of the coil (figure \ref{f.1V_cur_cooling}). As the current overcomes the critical current $I_c = 470$\,A (point B), localized heating triggers a transition to the normal state (figure \ref{f.1V_temp_cooling}), leading to a sudden increase in resistance and a sharp rise in temperature (point C), followed by a drop in current. Subsequently, the coil enters a self-oscillatory region (points D-F and beyond), where both current and maximum temperature exhibit oscillations (figure \ref{f.1V_cur_cooling} and \ref{f.1V_temp_cooling}). These oscillations result from a dynamic feedback mechanism: local heating causes sections of the coil to become resistive, reducing current; the reduced current lowers Joule heating, allowing the coil to cool; superconductivity is then partially restored, and current begins to rise again, repeating the cycle. This behavior reflects the coupling between superconducting transitions, inductive dynamics, and anisotropic heat transfer.

\begin{figure}[htp]
\subfloat[initial ramp]{
  {\includegraphics[trim=1.6cm 0 1.4cm 0,clip,height=4.1cm]{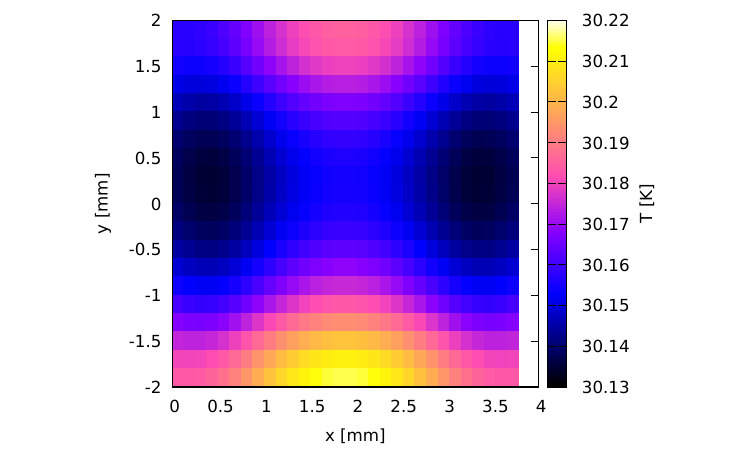}}
}
\subfloat[initial ramp]{
  {\includegraphics[trim=1.6cm 0 1.6cm 0,clip,height=4.1cm]{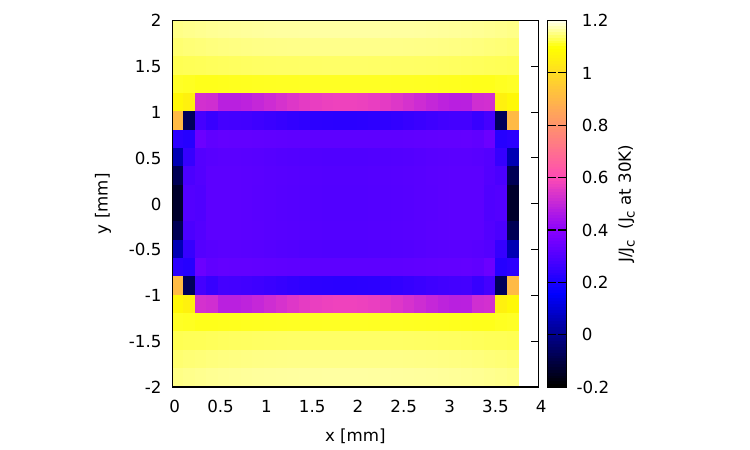}}
}
\subfloat[initial ramp]{
  {\includegraphics[trim=1.6cm 0 1.4cm 0,clip,height=4.1cm]{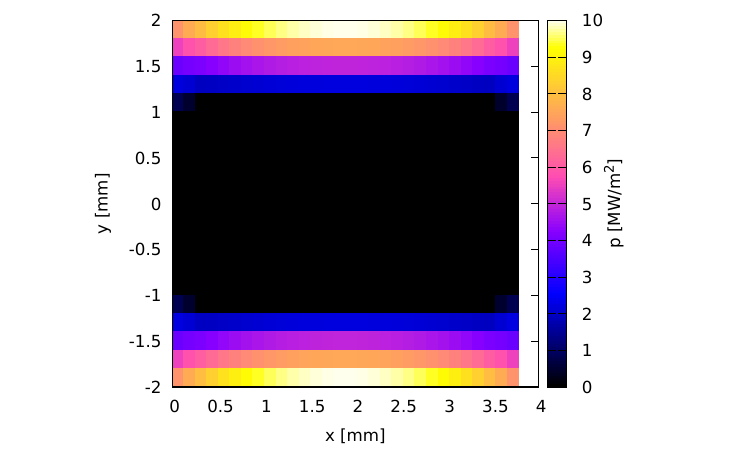}}
}

\hspace{0mm}
\subfloat[first current peak]{
  {\includegraphics[trim=1.6cm 0 1.4cm 0,clip,height=4.1cm]{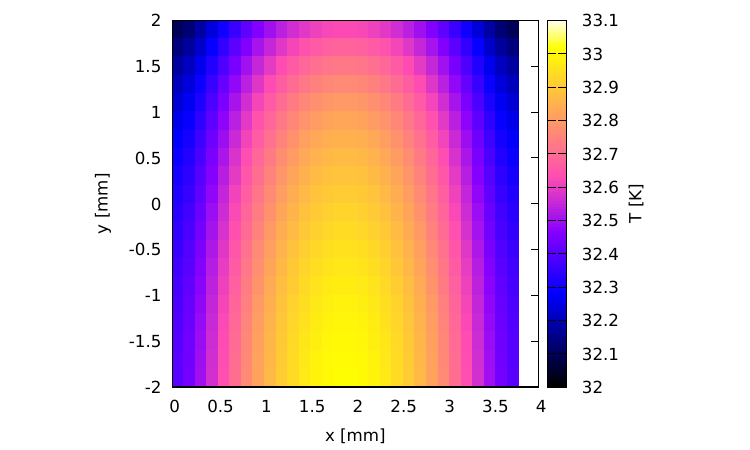}}
}
\subfloat[first current peak]{
  {\includegraphics[trim=1.6cm 0 1.6cm 0,clip,height=4.1cm]{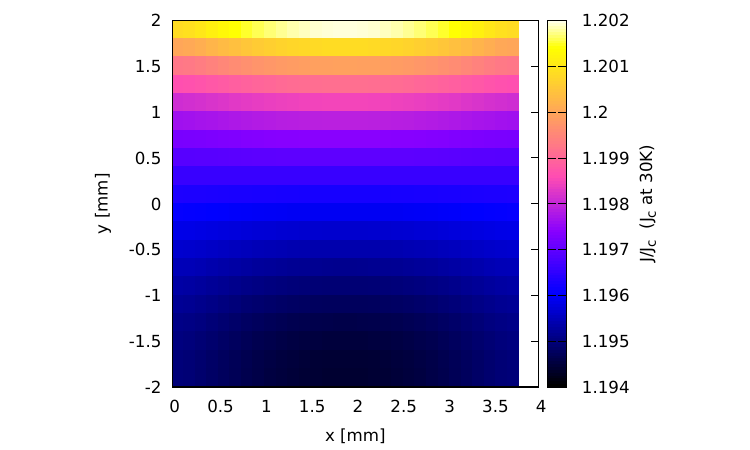}}
}
\subfloat[first current peak]{
  {\includegraphics[trim=1.6cm 0 1.4cm 0,clip,height=4.1cm]{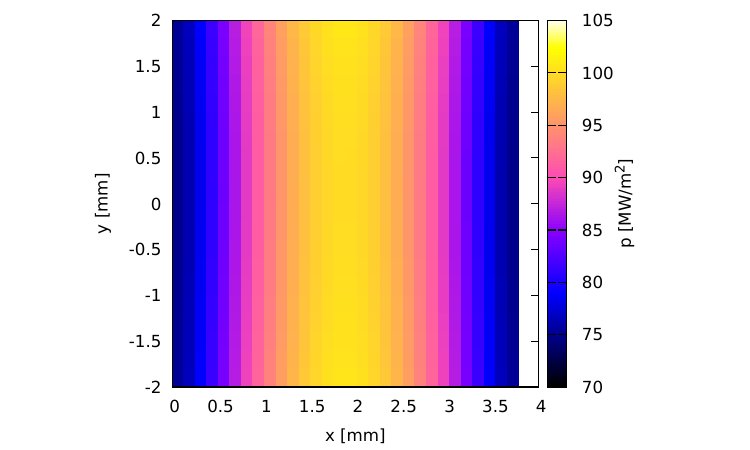}}
}

\hspace{0mm}
\subfloat[first drop]{
  {\includegraphics[trim=1.6cm 0 1.4cm 0,clip,height=4.1cm]{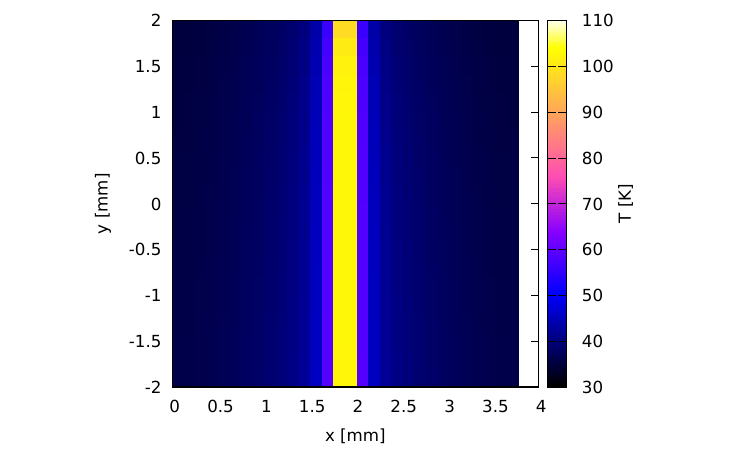}}
}
\subfloat[first drop]{
  {\includegraphics[trim=1.6cm 0 1.6cm 0,clip,height=4.1cm]{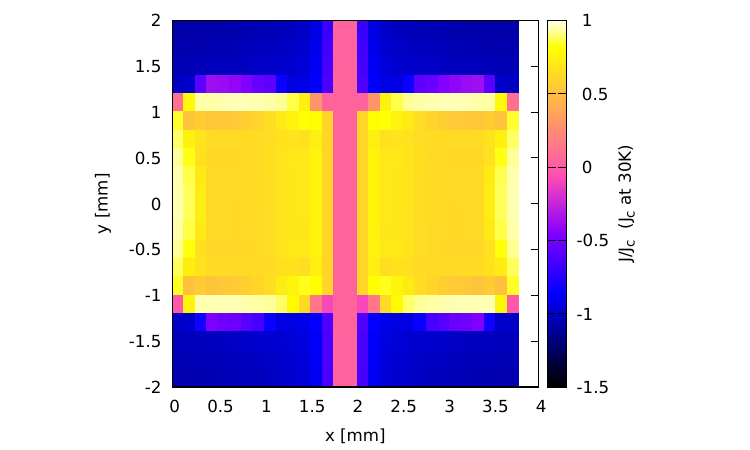}}
}
\subfloat[first drop]{
  {\includegraphics[trim=1.6cm 0 1.4cm 0,clip,height=4.1cm]{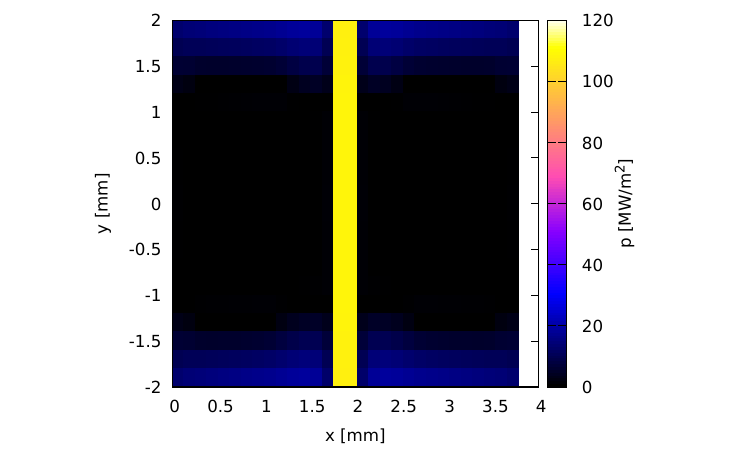}}
}

\hspace{0mm}
\subfloat[second current Peak]{
  {\includegraphics[trim=1.6cm 0 1.4cm 0,clip,height=4.1cm]{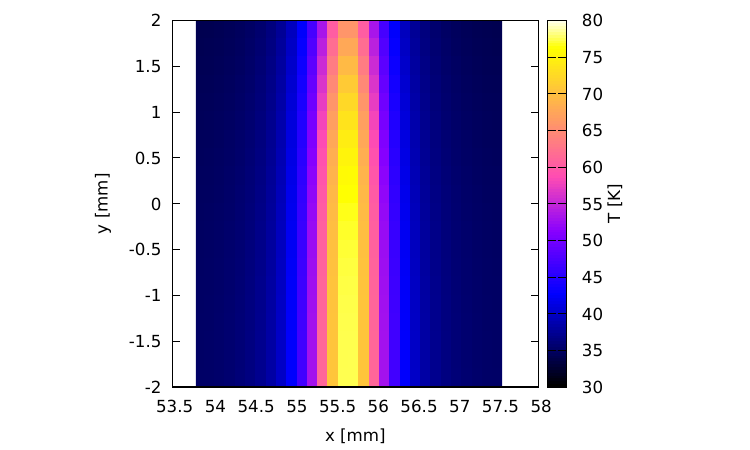}}
}
\subfloat[second current Peak]{
  {\includegraphics[trim=1.6cm 0 1.6cm 0,clip,height=4.1cm]{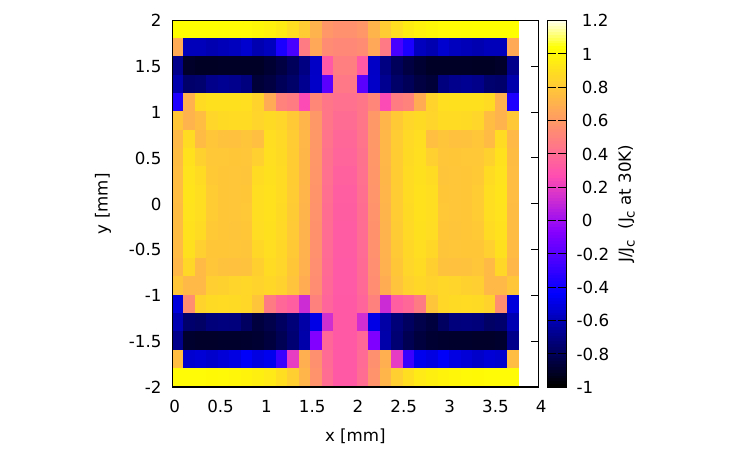}}
}
\subfloat[second current Peak]{
  {\includegraphics[trim=1.6cm 0 1.4cm 0,clip,height=4.1cm]{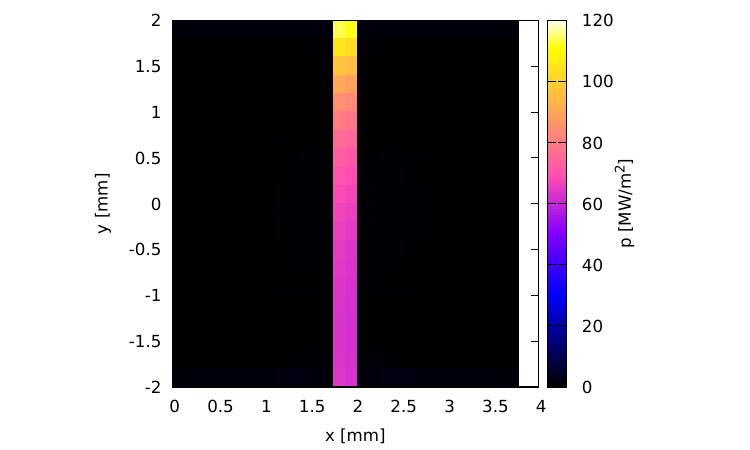}}
}

\caption{Coil cross-section showing maps of temperature (left), current density (middle), and power loss density (right) for top cooling at 1 V amplitude of DC voltage for different times: (a,b,c) at point A, (d,e,f) at point B, (g,h,i) at point C, and (j,k,l) at point D in figures \ref{f.1V_cur_cooling} and \ref{f.1V_temp_cooling}.}
\label{f.maps_1Vdc_cooling}
\end{figure}

Next, we analyze the temperature, normalized current density and power-loss distributions at crucial time steps in figure \ref{f.maps_1Vdc_cooling}. The behavior up to point C is the same as the adiabatic case (figures \ref{f.maps_1Vdc_cooling} (a-i)), except that the temperature close to the top surface is lower, thanks to conduction cooling. At the peak of current (point B), the current density close to the top is higher because the lower temperature causes higher local $J_c$. We further analyze the distributions at the second current peak (point D in figure \ref{f.1V_cur_cooling}). Between points C and D in figures \ref{f.1V_cur_cooling} and \ref{f.1V_temp_cooling}, the maximum temperature decreases because of heat diffusion to their neighboring turns (figure \ref{f.maps_1Vdc_cooling}(j)). This causes that the maximum temperature, which is at the central turns, becomes lower than $T_c$. Then, the whole coil becomes superconducting, and hence the current increases until it overcomes the coil critical current for the given temperature distribution (point D in figure \ref{f.1V_cur_cooling} and figure \ref{f.maps_1Vdc_cooling}(k)). This causes local heat at the central turns (figure \ref{f.maps_1Vdc_cooling}(l)), which later causes another temperature increase between points D and E in figure \ref{f.1V_temp_cooling} and its corresponding current decrease when the temperature overcomes the critical temperature. All this process causes self-oscillations in current and maximum temperature.


\subsection{Higher thermal conducting inter-layer}

In this study, we examine the revised geometry incorporating tape-to-tape insulation with enhanced thermal conductivity, which is set to that of stainless steel for testing purposes. The metallic substrate thickness is reduced from 100 $\mu$m to 80 $\mu$m to reflect a more compact and thermally responsive design. We analyze the 1 V case and consider both adiabatic and conduction cooling conditions. We chose these parameters in order to observe higher oscillations.

\subsubsection{Adiabatic Condition}

\begin{figure}[htp]
\centering
\includegraphics[trim=0 0 0 0,clip,width=10 cm]{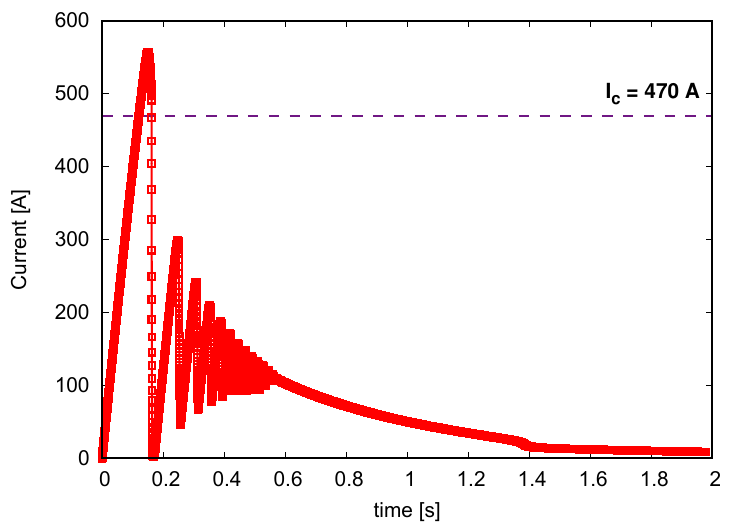}
\\
\includegraphics[trim=0 0 0.36cm 0,clip,width=10 cm]{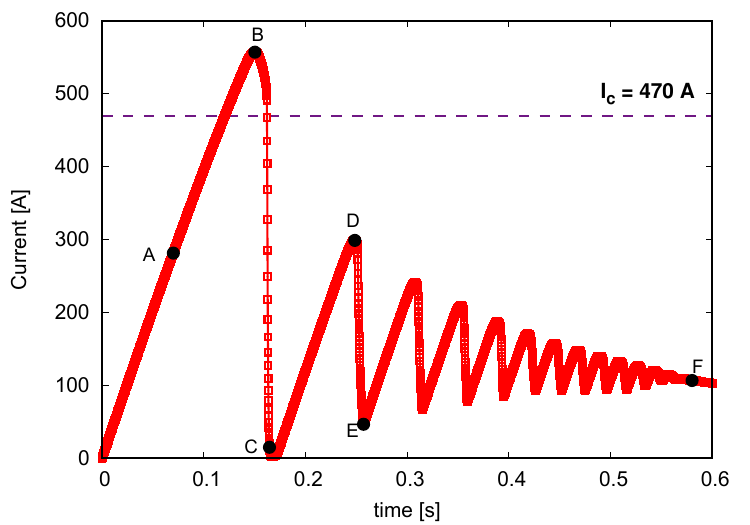}
\caption{(a) Current in racetrack coil for 1 V DC input with no heat exchange with the cryogenic liquid. (b) Zoomed sketch of the above results from $0 - 0.6$ s.}
\label{f.1V_cur}
\end{figure}

\begin{figure}[htp]
\centering
\includegraphics[trim=0 0 0 0,clip,width=10 cm]{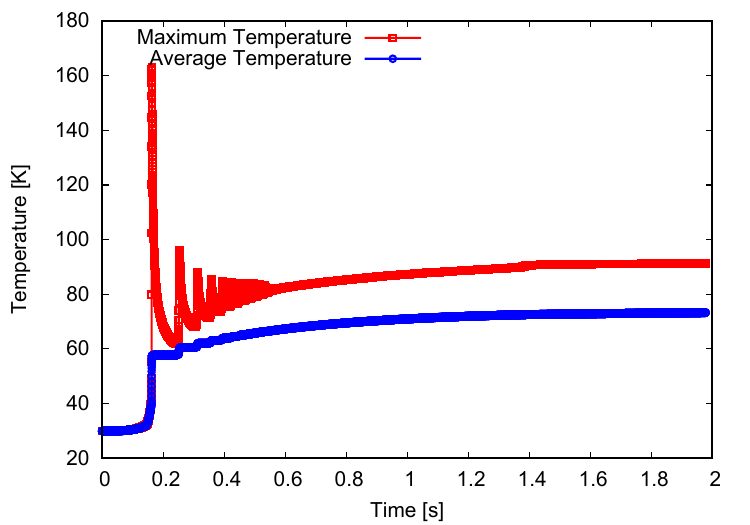}
\\
\includegraphics[trim=0 0 0 0,clip,width=10 cm]{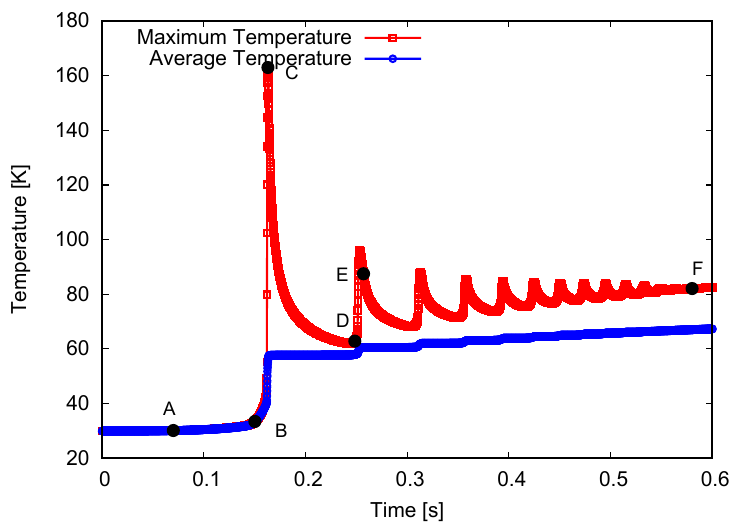}
\caption{ (a) Temperature rise in racetrack coil for 1 V DC input with no heat exchange with the cryogenic liquid. (b) Zoomed sketch of the above results from $0 - 0.6$ s.}
\label{f.1V_temp}
\end{figure}

Initially, the current increases sharply and then drops, similar to previous cases, but subsequently begins oscillating with gradually decreasing amplitude, ultimately approaching to zero (figure \ref{f.1V_cur}). This fluctuating current behavior leads to temperature spikes, each almost aligning with the current oscillations (figure \ref{f.1V_temp}). Notably, while the maximum temperature experiences fluctuations, the overall average temperature rises steadily, although the latter remains below the critical threshold. In addition, the oscillations in the maximum temperature continue till $0.6$ s and this temperature becomes smooth afterwards (figure \ref{f.1V_temp}).

\begin{figure}[htp]
\subfloat[initial ramp]{
  {\includegraphics[trim=1.6cm 0 1.4cm 0,clip,height=4.1cm]{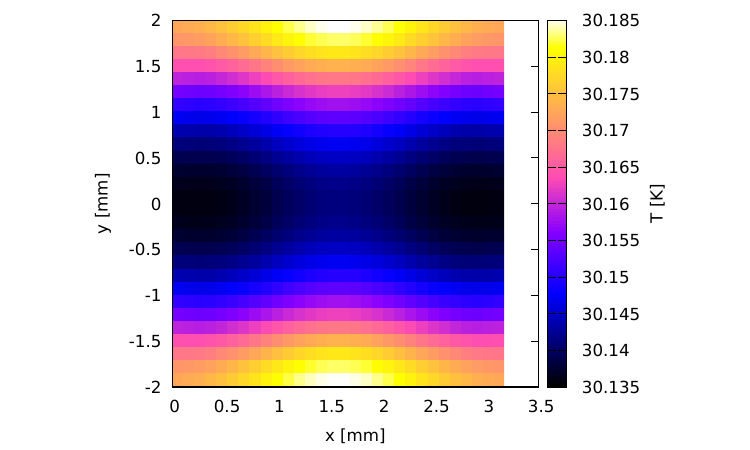}}
}
\subfloat[initial ramp]{
  {\includegraphics[trim=1.6cm 0 1.9cm 0,clip,height=4.1cm]{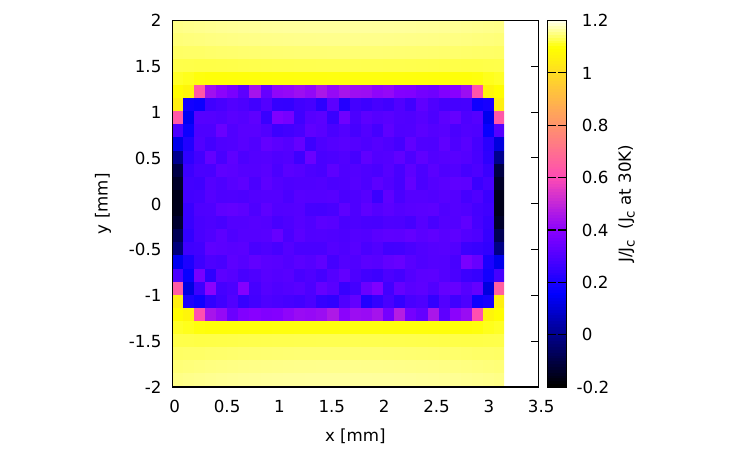}}
}
\subfloat[initial ramp]{
  {\includegraphics[trim=1.6cm 0 1.4cm 0,clip,height=4.1cm]{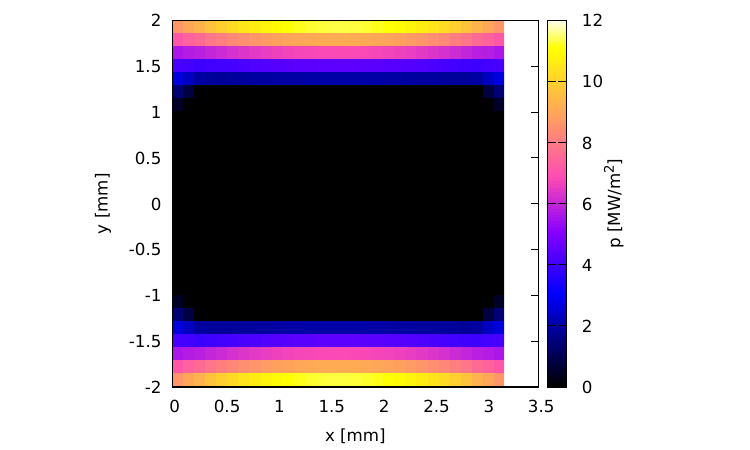}}
}

\hspace{0mm}
\subfloat[first current peak]{
  {\includegraphics[trim=1.6cm 0 1.4cm 0,clip,height=4.1cm]{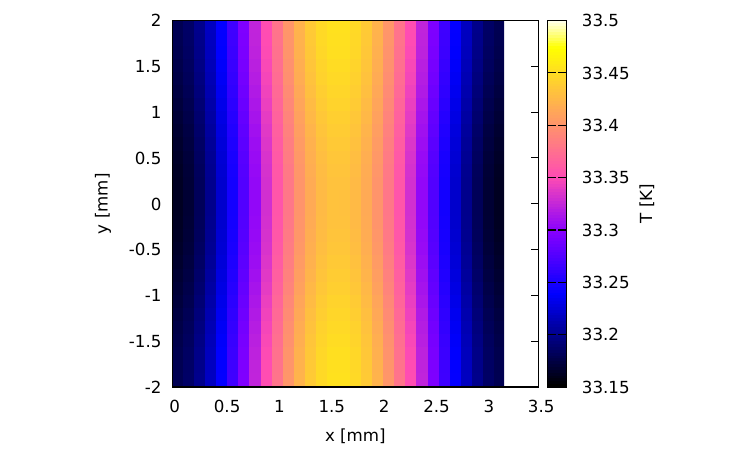}}
}
\subfloat[first current peak]{
  {\includegraphics[trim=1.6cm 0 1.6cm 0,clip,height=4.1cm]{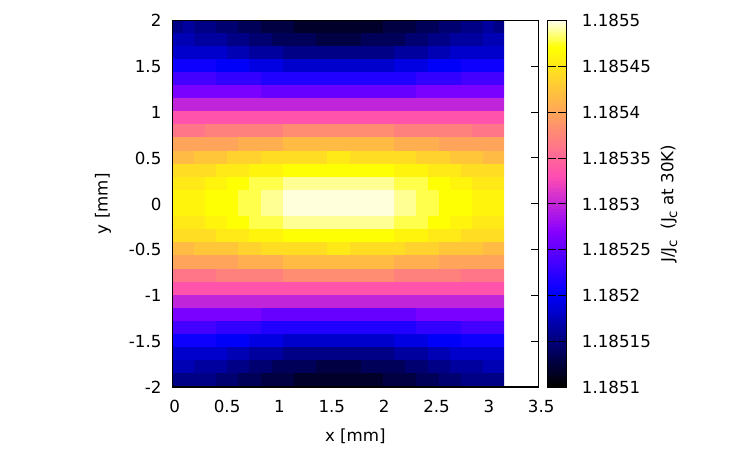}}
}
\subfloat[first current peak]{
  {\includegraphics[trim=1.6cm 0 1.4cm 0,clip,height=4.1cm]{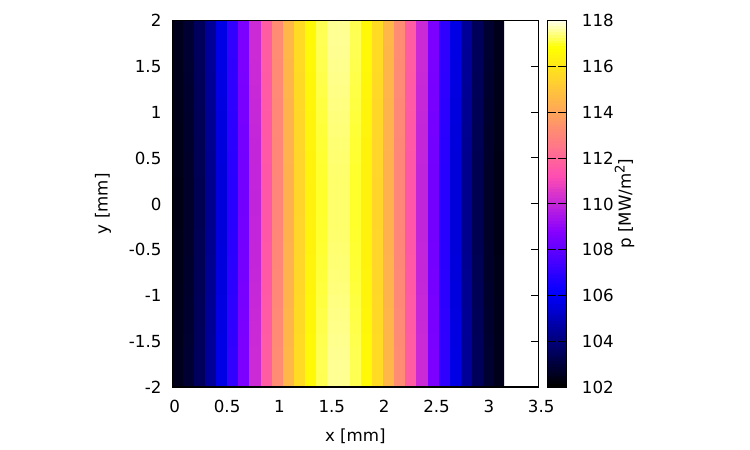}}
}

\hspace{0mm}
\subfloat[first drop]{
  {\includegraphics[trim=1.6cm 0 1.4cm 0,clip,height=4.1cm]{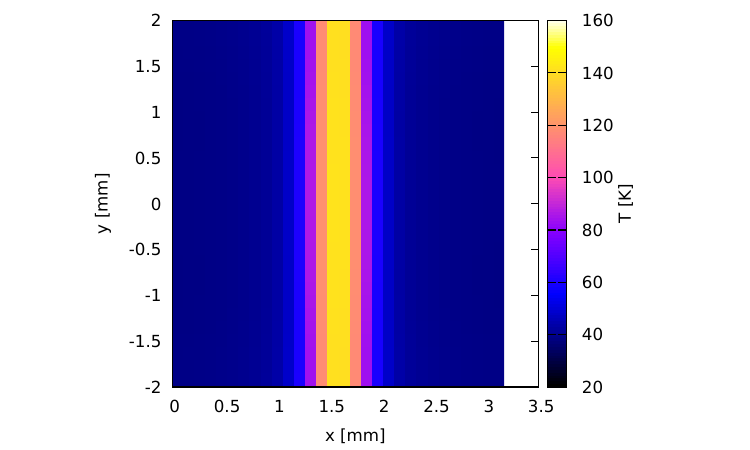}}
}
\subfloat[first drop]{
  {\includegraphics[trim=1.6cm 0 1.9cm 0,clip,height=4.1cm]{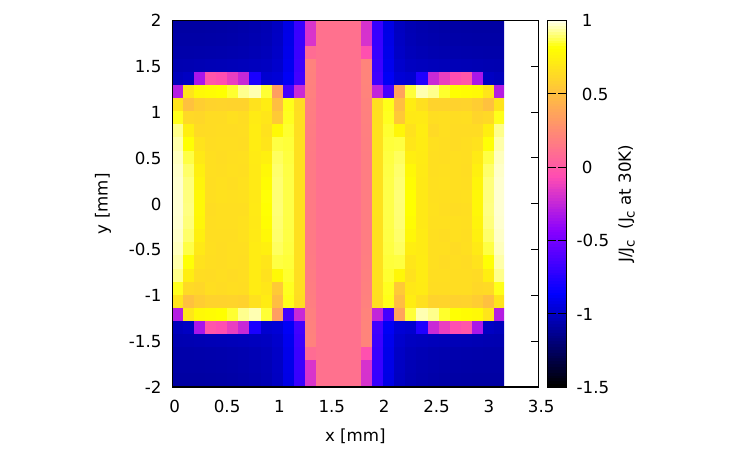}}
}
\subfloat[first drop]{
  {\includegraphics[trim=1.6cm 0 1.4cm 0,clip,height=4.1cm]{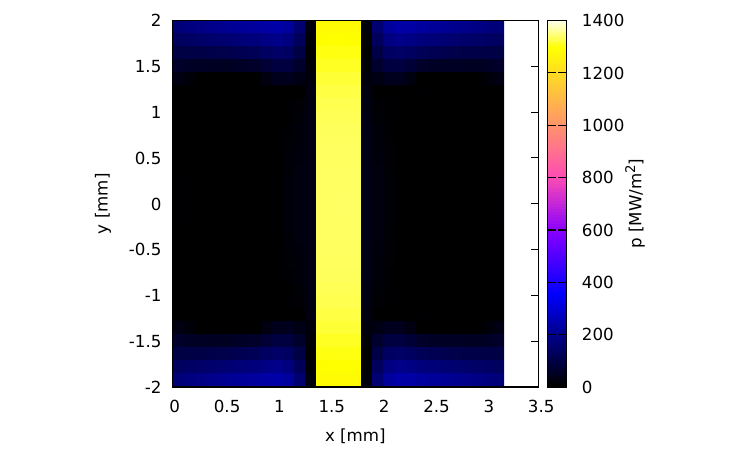}}
}

\hspace{0mm}
\subfloat[second current peak]{
  {\includegraphics[trim=1.6cm 0 1.4cm 0,clip,height=4.1cm]{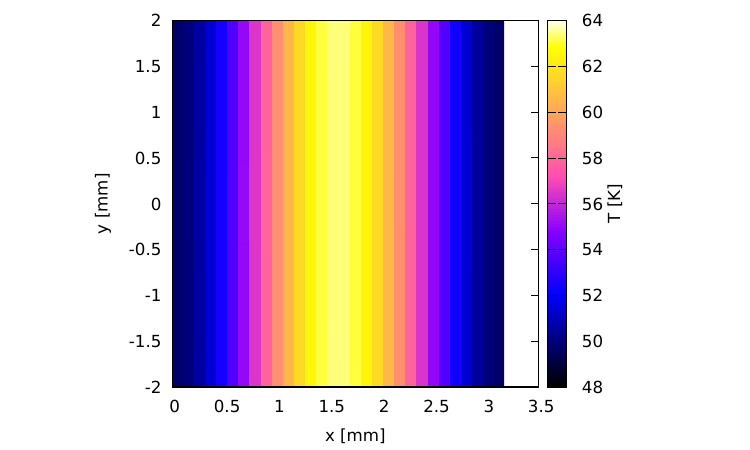}}
}
\subfloat[second current peak]{
  {\includegraphics[trim=1.6cm 0 1.9cm 0,clip,height=4.1cm]{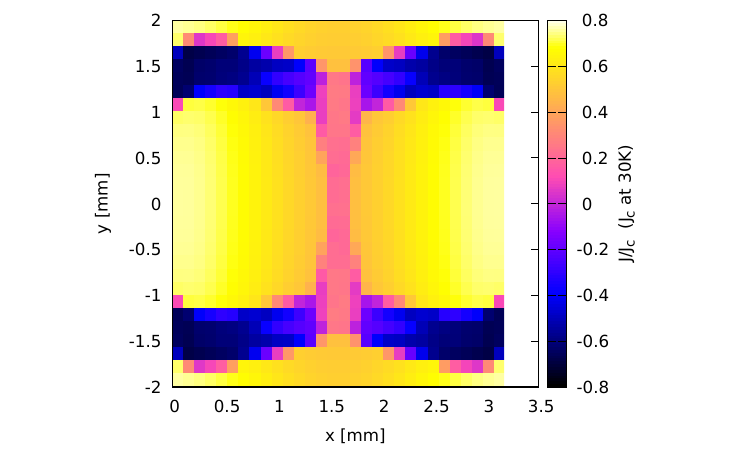}}
}
\subfloat[second current peak]{
  {\includegraphics[trim=1.6cm 0 1.4cm 0,clip,height=4.1cm]{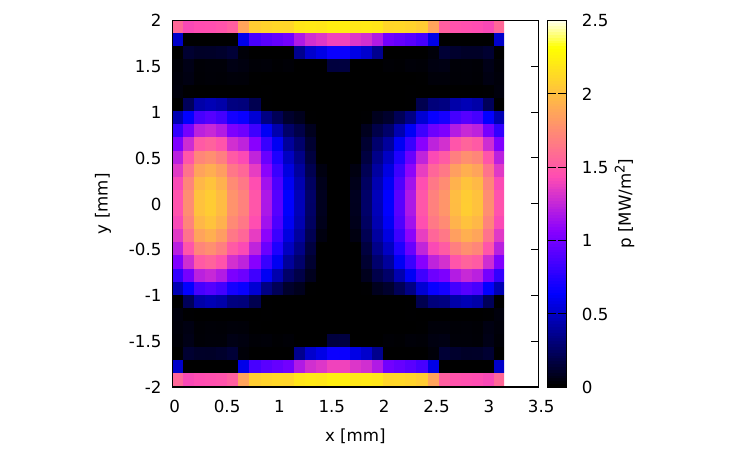}}
}

\caption{Coil cross-section showing maps of temperature (left), current density (middle), and power loss density (right) for adiabatic conditions at 1 V amplitude of DC voltage for different times: (a,b,c) at point A (d,e,f) at point B, (g,h,i) at point C, and (j,k,l) at point D in figures \ref{f.1V_cur} and \ref{f.1V_temp}.}
\label{f.maps_silver_1Vdc}
\end{figure}

Next we inspect the temperature, current density, and power loss density at specific time steps (see figure \ref{f.maps_silver_1Vdc}). The behavior during the first current peak is qualitatively the same as for electric insulation with low thermal conductivity like polyimide (points A, B, C in figures \ref{f.1V_cur} and \ref{f.1V_temp}, which correspond to figures \ref{f.maps_silver_1Vdc}(a,b,c), \ref{f.maps_silver_1Vdc}(d,e,f), \ref{f.maps_silver_1Vdc}(g,h,i), respectively). The difference is that after the first drop in current (and peak in maximum $T$), the current steadily increases to higher values (region between C and D in figure \ref{f.1V_cur}). This is because after the first drop, there are very few turns with $T>T_c$. Then, the thermal diffusion that follows reduces $T$ below $T_c$ for all turns, which causes low resistance, and hence high current but low power loss (figures \ref{f.maps_silver_1Vdc}(j,k,l)). 

\begin{figure}[htp]
\subfloat[initial ramp]{
  {\includegraphics[trim=1.6cm 0 1.4cm 0,clip,height=4.1cm]{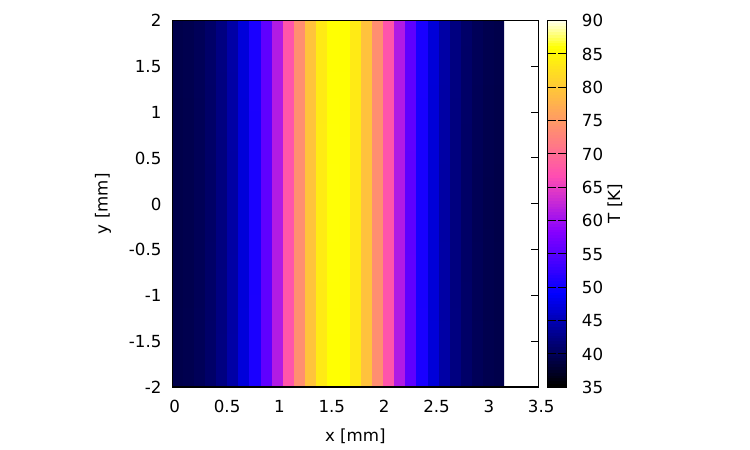}}
}
\subfloat[initial ramp]{
  {\includegraphics[trim=1.6cm 0 1.4cm 0,clip,height=4.1cm]{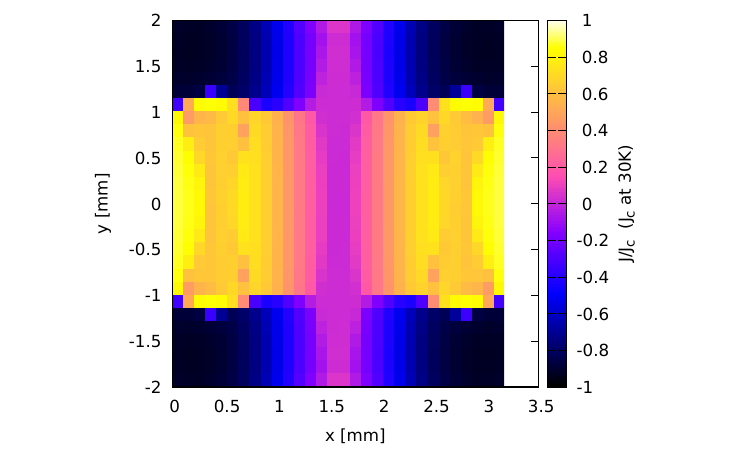}}
}
\subfloat[initial ramp]{
  {\includegraphics[trim=1.6cm 0 1.4cm 0,clip,height=4.1cm]{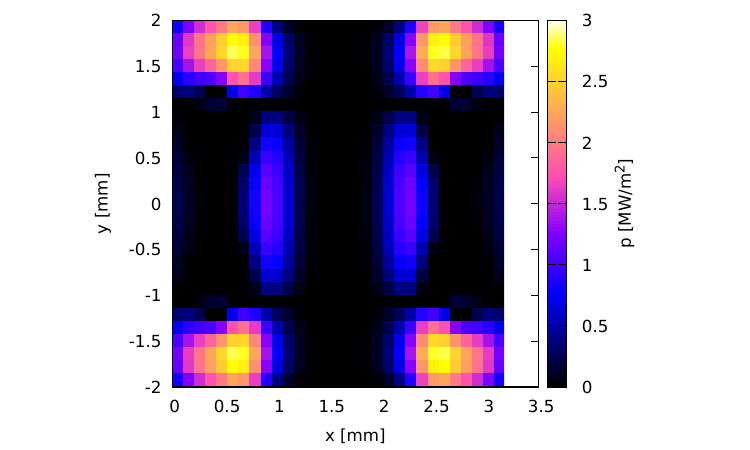}}
}

\hspace{0mm}
\subfloat[first current peak]{
  {\includegraphics[trim=1.6cm 0 1.4cm 0,clip,height=4.1cm]{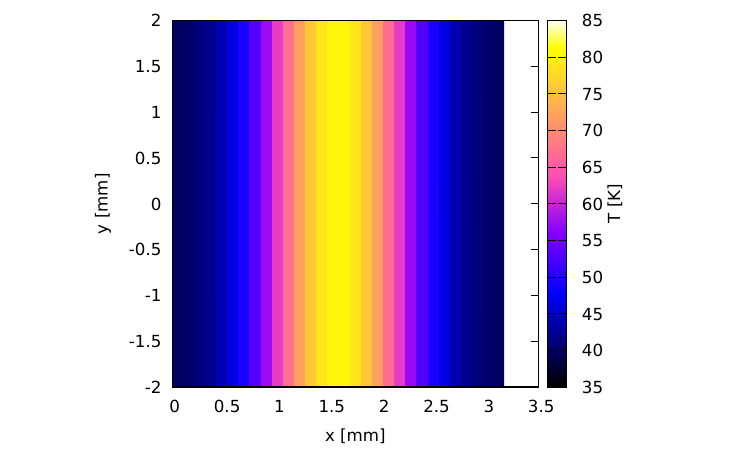}}
}
\subfloat[first current peak]{
  {\includegraphics[trim=1.6cm 0 1.4cm 0,clip,height=4.1cm]{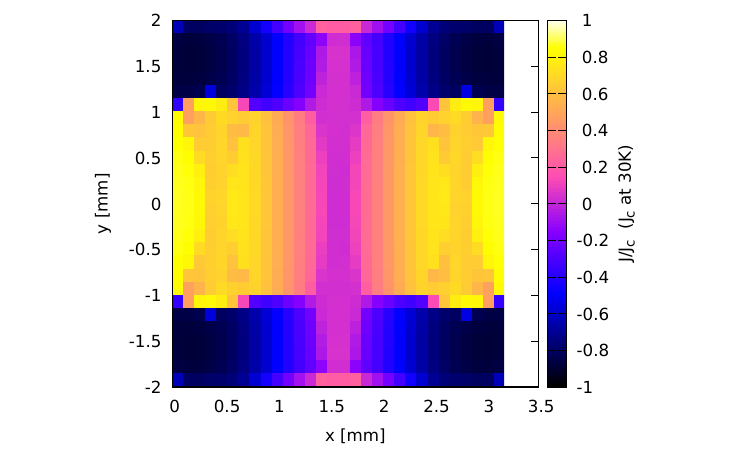}}
}
\subfloat[first current peak]{
  {\includegraphics[trim=1.6cm 0 1.4cm 0,clip,height=4.1cm]{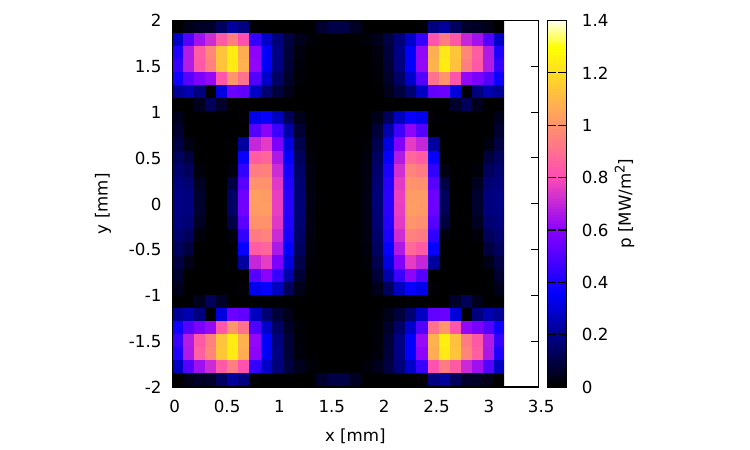}}
}
\caption{Coil cross-section showing maps of temperature (left), current density (middle), and power loss density (right) for adiabatic conditions at 1 V amplitude of DC voltage for different times: (a,b,c) at point E and (d,e,f) at point F in figures \ref{f.1V_cur} and \ref{f.1V_temp}.}
\label{f.maps_silver_1Vdc_1}
\end{figure}

The remaining dissipation causes temperature increase until it becomes close to the critical current for a few central turns (figure \ref{f.maps_silver_1Vdc_1}(a,b,c)). This causes an increase in the coil resistance, and hence a decrease in current (point E in figure \ref{f.1V_cur}). Again, the low current causes low AC loss, and hence the temperature in the central turns decreases due to thermal diffusion. This will again cause yet another decrease in coil resistance and consequent increase in current. Each oscillation has lower amplitude because the difference between the minimum and maximum temperature across the coil cross-section decreases, and hence diffusion after the current drop is less efficient. We further analyze the temperature and current density distribution during the gradual temperature increase when there are no oscillations. At this stage (point F in figure \ref{f.1V_cur}), the temperature rises slowly, allowing sufficient time for heat to diffuse into the neighboring turns. This effectively limits localized temperature increase, which was causing previous oscillations.


\subsubsection{Conduction Cooling}

\begin{figure}[htp]
\centering
\includegraphics[trim=0 0 0 0,clip,width=9 cm]{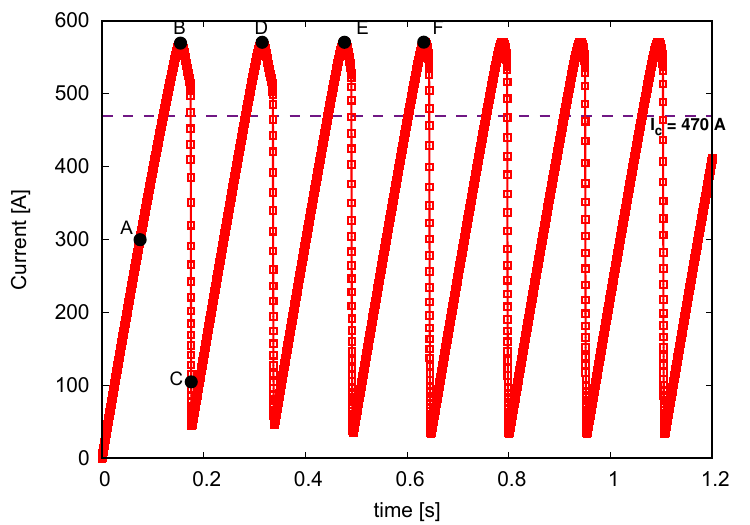}
\caption{Total current in racetrack coil for 1 V DC input with with cooling from the top at $30$ K.}
\label{f.1Vdc_cur_cooling02}
\end{figure}

\begin{figure}[htp]
\centering
\includegraphics[trim=0 0 0 0,clip,width=9 cm]{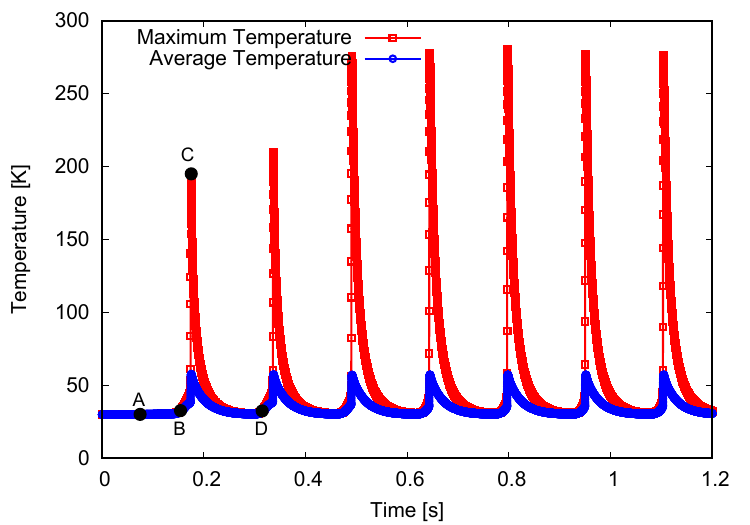}
\caption{Temperature rise in racetrack coil for 1 V DC input with cooling from the top at $30$ K.}
\label{f.1Vdc_temp_cooling01}
\end{figure}

Now, the racetrack coil exhibits completely periodic thermal and current behaviors (figure \ref{f.1Vdc_cur_cooling02} and \ref{f.1Vdc_temp_cooling01}). After the peak of the current (figure \ref{f.1Vdc_cur_cooling02} point B), we observe a sharp rise in temperature due to non-linear Joule heating. This is followed by a drop in maximum temperature and a subsequent temperature rise, creating a cyclic pattern (figure \ref{f.1Vdc_temp_cooling01}). The cooling mechanism from the top surface plays a crucial role in avoiding thermal damage of the superconductor and also causes perfectly periodic oscillations of temperature and current after a transient, which extends to the third temperature peak. However, at certain point a circuit breaker should act in order to avoid mechanical fatigue due to cyclic thermal stress. Indeed, temperature oscillations have high amplitude, where the maximum temperature ranges from around 30 to 280 K at a rate of roughly 6 times per second.

\begin{figure}[htp]
\subfloat[initial ramp]{
  {\includegraphics[trim=1.6cm 0 1.4cm 0,clip,height=4.1cm]{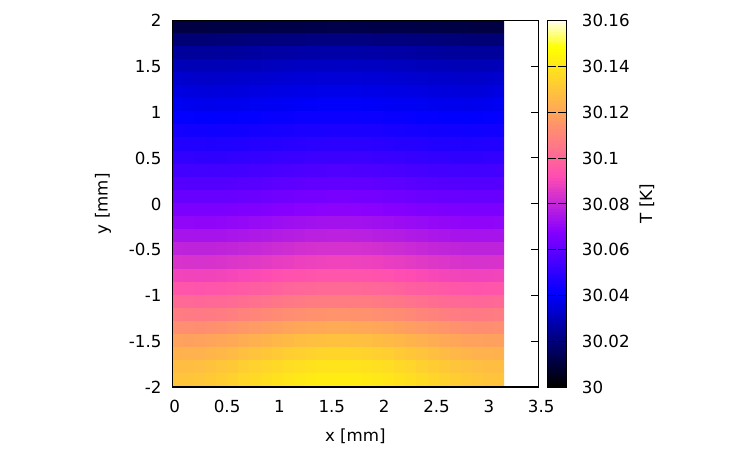}}
}
\subfloat[initial ramp]{
  {\includegraphics[trim=1.6cm 0 1.6cm 0,clip,height=4.1cm]{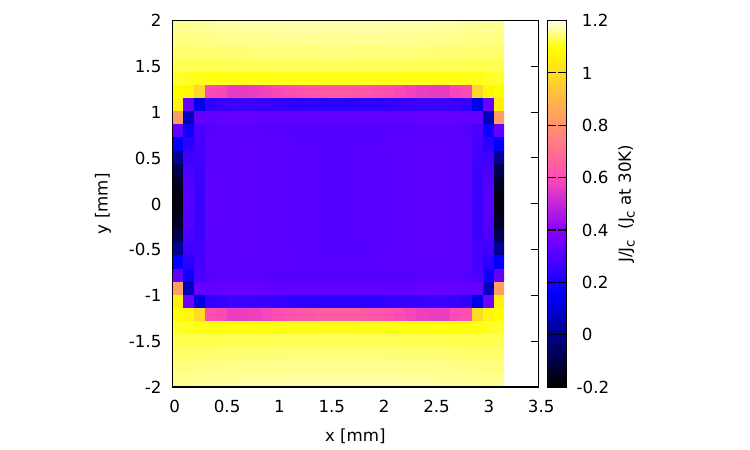}}
}
\subfloat[initial ramp]{
  {\includegraphics[trim=1.6cm 0 1.4cm 0,clip,height=4.1cm]{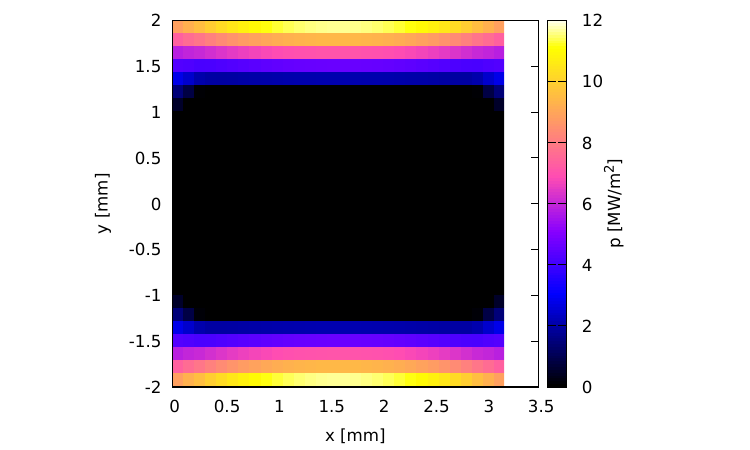}}
}

\hspace{0mm}
\subfloat[first current peak]{
  {\includegraphics[trim=1.6cm 0 1.4cm 0,clip,height=4.1cm]{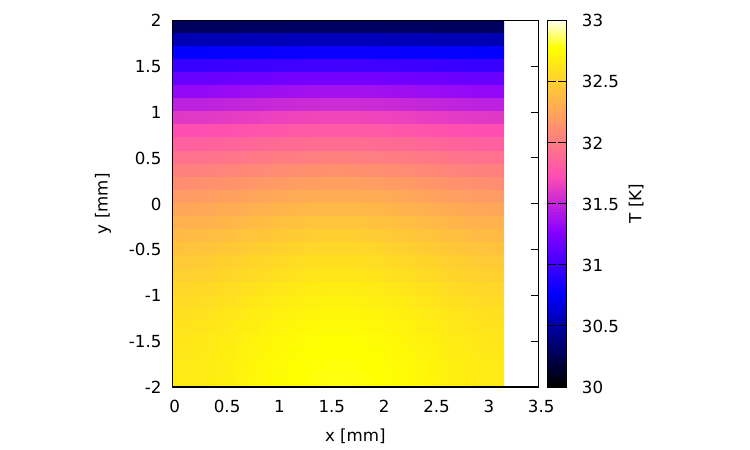}}
}
\subfloat[first current peak]{
  {\includegraphics[trim=1.6cm 0 1.6cm 0,clip,height=4.1cm]{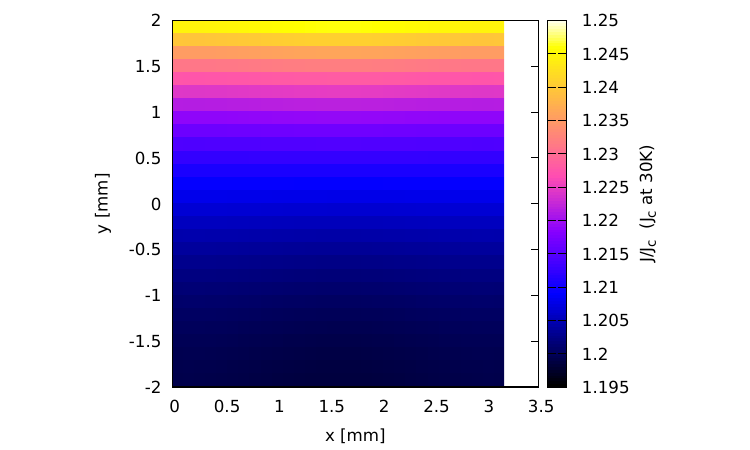}}
}
\subfloat[first current peak]{
  {\includegraphics[trim=1.6cm 0 1.4cm 0,clip,height=4.1cm]{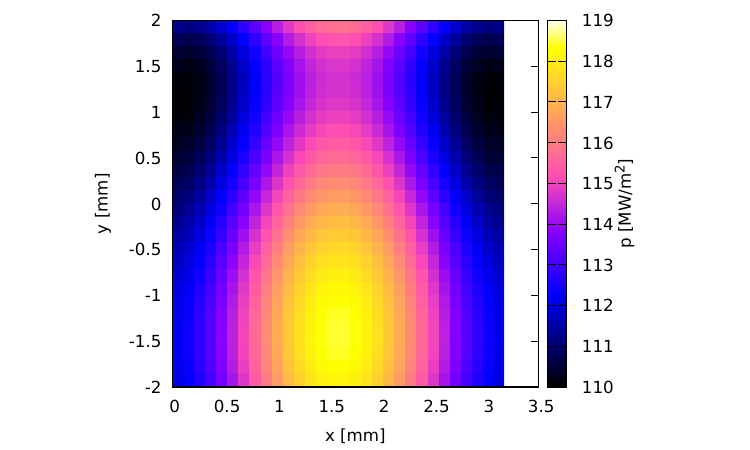}}
}

\hspace{0mm}
\subfloat[first drop]{
  {\includegraphics[trim=1.6cm 0 1.4cm 0,clip,height=4.1cm]{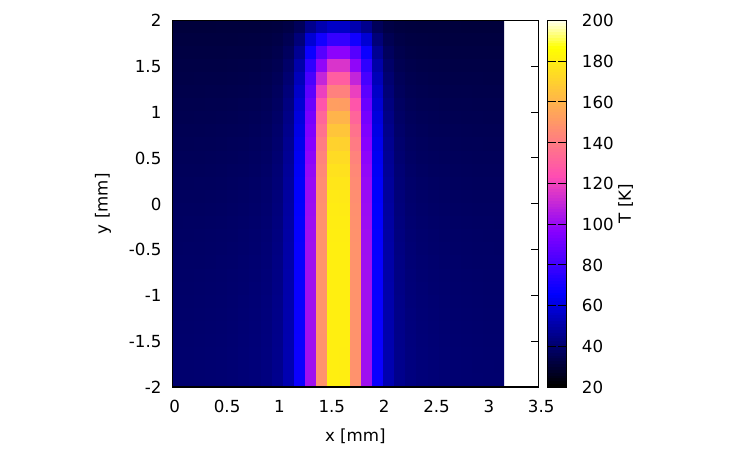}}
}
\subfloat[first drop]{
  {\includegraphics[trim=1.6cm 0 1.6cm 0,clip,height=4.1cm]{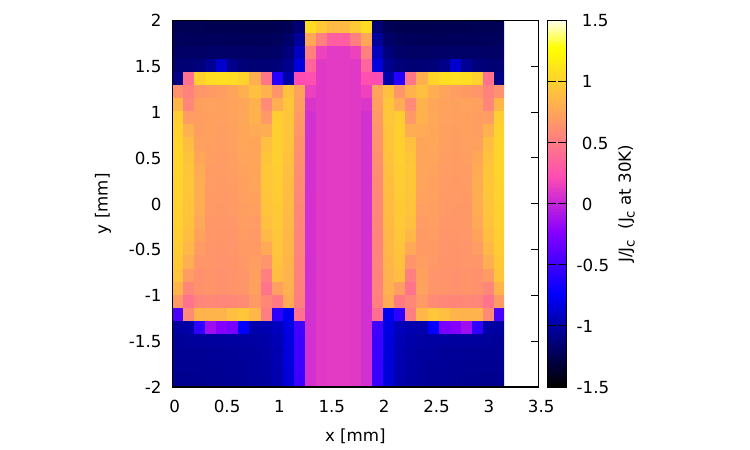}}
}
\subfloat[first drop]{
  {\includegraphics[trim=1.6cm 0 1.4cm 0,clip,height=4.1cm]{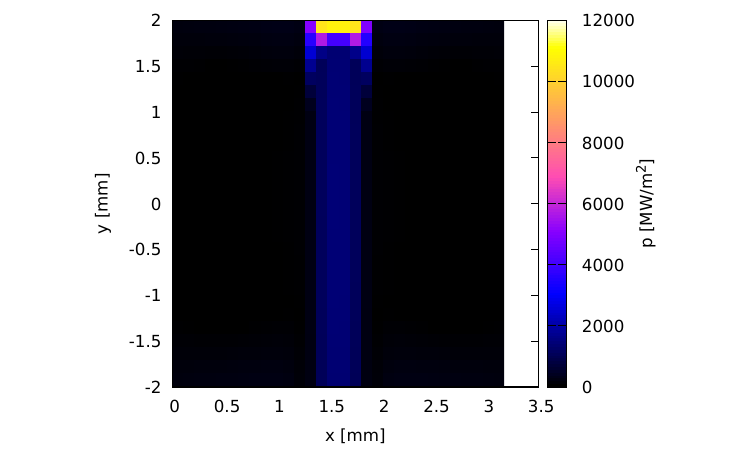}}
}

\hspace{0mm}
\subfloat[second current Peak]{
  {\includegraphics[trim=1.6cm 0 1.4cm 0,clip,height=4.1cm]{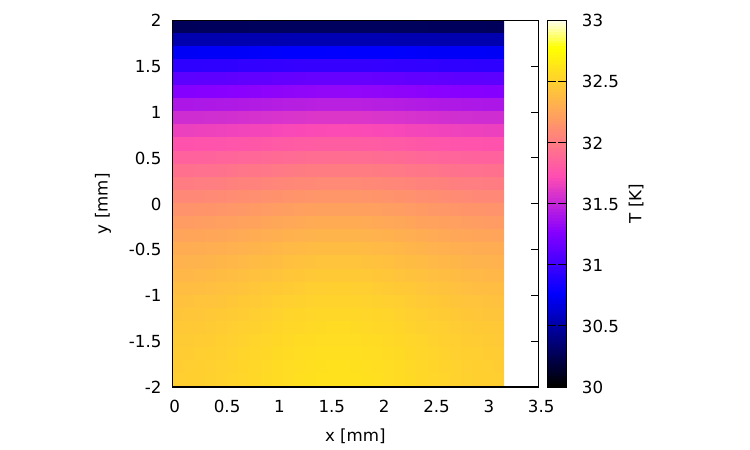}}
}
\subfloat[second current Peak]{
  {\includegraphics[trim=1.6cm 0 1.6cm 0,clip,height=4.1cm]{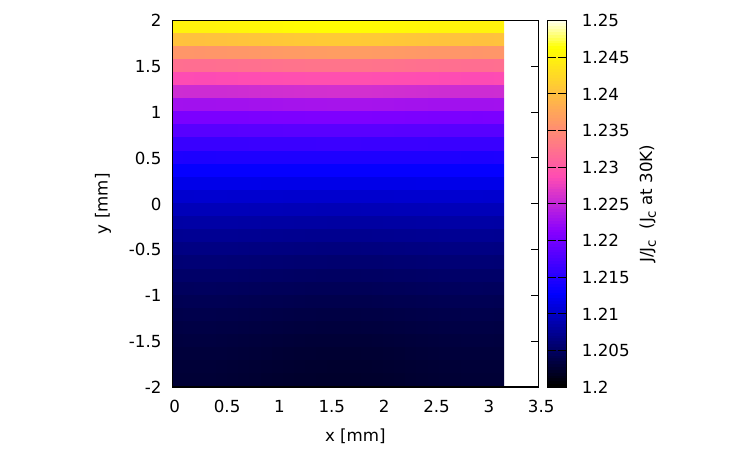}}
}
\subfloat[second current Peak]{
  {\includegraphics[trim=1.6cm 0 1.4cm 0,clip,height=4.1cm]{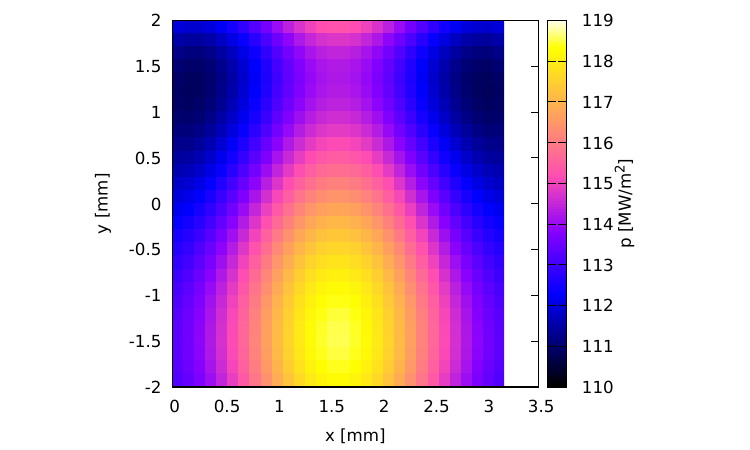}}
}

\caption{Coil cross-section showing maps of temperature (left), current density (middle), and power loss density (right) for top cooling at 1 V amplitude of DC voltage for different times: (a,b,c) at point A, (d,e,f) at point B, (g,h,i) at point C, and (j,k,l) at point D in figures \ref{f.1Vdc_cur_cooling02} and \ref{f.1Vdc_temp_cooling01}.}
\label{f.maps_1Vdc_cooling03}
\end{figure}

Next, we conduct a detailed analysis of the behavior by examining the distributions of current, temperature, and power dissipation. The current density and temperature distributions at the initial ramp again show the presence of screening currents (figure \ref{f.maps_1Vdc_cooling03}(b)). During the initial current ramp, the temperature on the top side of the coil remains practically unchanged due to conduction cooling (figure \ref{f.maps_1Vdc_cooling03}(a)). The distribution of current density and the resulting power loss is almost the same as without cooling (figures \ref{f.maps_1Vdc_cooling03}(b,c)). At the current peak (point B in figure \ref{f.1Vdc_cur_cooling02} and \ref{f.1Vdc_temp_cooling01}), we observe temperature rise, the effect of cooling on the top surface of the coil, current densities above $J_c$ of 30 K at the whole cross-section, and roughly uniform power loss (see figure \ref{f.maps_1Vdc_cooling03}(d,e,f)). We also observe the effect of cooling on the top surface at the first drop of current (point C in figures \ref{f.1Vdc_cur_cooling02} and \ref{f.1Vdc_temp_cooling01}), where the temperature at the central turns is above the critical temperature ($T_c$ = 92 K), while the top surface portion of the coil is colder (figure \ref{f.maps_1Vdc_cooling03}(g)). The current drop in the central turns is significant because their temperature is above above $T_c$ in most of the tape, and hence they are in normal state (figure \ref{f.maps_1Vdc_cooling03}(g,h)) except a small portion at the top, which gets saturated with $J$ around $J_c$ at the local temperature. At figure \ref{f.maps_1Vdc_cooling03}(i), the power density accumulates at the top of the central turns. This is because $J$ is above the local $J_c$, which generates relatively high power heat, $p=J\cdot E(J)$, due to the superconductor power-law ${\bf E}({\bf J})$ relation. 


\section{Conclusion}

In this article, we investigated the thermal quench dynamics of a REBCO racetrack coil with parameters typical for aviation motors under conduction cooling at 30 K. To achieve realistic and accurate results, we implemented temperature-dependent material properties in our coupled electromagnetic and electrothermal models. This approach allowed us to capture the complex interplay between thermal quench, heat generation, and power loss.

Our findings highlight that the thermal quench behavior strongly depends on the fault voltage. At higher voltages, such as 1000 V DC, rapid quench propagation occurs due to excessive Joule heating and transition to normal state, resulting in a fast temperature rise beyond the critical temperature ($T_c$). Conversely, at lower voltages, particularly at 100 V and below, conduction cooling plays a crucial role in stabilizing the coil temperature by efficiently extracting heat and avoiding high temperatures that can damage the superconductor.  
At 1V DC, we observe a different response. The coil shows self-oscillations in both temperature and current. In order to achieve oscillations it is necessary to have: dissipation, cooling, and thermal diffusion between turns. These oscillations increase when increasing the thermal conductivity of the electric isolation between turns and reducing the substrate thickness. This is because in this way we increase thermal diffusivity between turns, and hence thermal diffusion effects are enhanced.

This study provides interesting insights into the quench behavior of conduction-cooled REBCO coils, offering valuable guidelines for the design and protection of superconducting aviation motors. The developed model serves as an effective tool for optimizing key design parameters, such as copper thickness, to enhance the thermal and electrical stability of superconducting systems. A critical technical implication of this study is the required response time for circuit breakers under different voltage conditions in an adiabatic environment. For higher voltages, such as $1000$ V, quench occurs rapidly, requiring ultra-fast circuit breakers with response times in the millisecond range to prevent excessive thermal runaway. In contrast, at lower voltages like $100$ V or below, conduction cooling plays a stabilizing role, even without circuit breakers. However, circuit breakers with integrated cooling may still be beneficial to improve system reliability under prolonged fault conditions.

\section*{Acknowledgement}
We acknowledge the financial support of \R{the EU NextGenerationEU through the Recovery and Resilience Plan for Slovakia under the project No. 09I04-03-V02-00039 and} the Slovak Republic from Projects APVV-19-0536 and VEGA 2/0098/24.

\end{document}